\documentclass[aps,pra,twocolumn,superscriptaddress,floatfix]{revtex4-2}

\usepackage{amssymb,amsmath}
\usepackage{graphicx}
\usepackage{dcolumn}
\usepackage{bm}
\usepackage[dvipsnames]{xcolor}
\usepackage{hyperref}
\hypersetup{colorlinks=false,
            pdfborder={0 0 0}}
\usepackage{environ}
\usepackage{braket}

\newcommand{\avg}[1]{\langle#1 \rangle}
\newcommand{\tr}[1]{\text{Tr}(#1)}

\newcommand{\UDSi}{5.5(1.0)}

\newcommand{\UDSv}{72(11)}
\newcommand{\TDSvi}{0.444(19)}
\newcommand{\UDSvi}{11.46(31)}

\newcommand{\UDSvii}{11.46(31)}

\newcommand{\UDSx}{5.32(21)}
\newcommand{\TDSxi}{0.74(13)}
\newcommand{\UDSxi}{5.32(21)}

\newcommand{\TDSxiii}{0.40(4)}
\newcommand{\UDSxiii}{26.5(2.6)}

\makeatletter
\newwrite\remember@figures
\AtBeginDocument{\InputIfFileExists{\jobname.dft}{}{}\immediate\openout\remember@figures=\jobname.dft}
\AtEndDocument{\immediate\closeout\remember@figures}
\NewEnviron{dfigure}[1]{%
    \immediate\write\remember@figures{%
        \noexpand\rememberfigure{#1}{\unexpanded\expandafter{\BODY}}%
    }%
}
\NewEnviron{dfigure*}[1]{%
    \immediate\write\remember@figures{%
        \noexpand\rememberfiguretc{#1}{\unexpanded\expandafter{\BODY}}%
    }%
}
\newcommand{\placefigure}[2][tp]{%
    \csname remembered@figure@#2\endcsname{#1}%
}
\newcommand{\rememberfigure}[2]{%
    \global\@namedef{remembered@figure@#1}##1{%
        \begin{figure}[##1]#2\end{figure}%
    }%
}
\newcommand{\rememberfiguretc}[2]{%
    \global\@namedef{remembered@figure@#1}##1{%
        \begin{figure*}[##1]#2\end{figure*}%
    }%
}
\makeatother

\begin{document}

\title{Observation of Nagaoka Polarons in a Fermi-Hubbard Quantum Simulator}

\newcommand{\harvard}{Department of Physics, Harvard University, Cambridge, MA 02138, USA}
\newcommand{\sjsu}{Department of Physics and Astronomy, San Jos\'e State University, San Jos\'e, CA 95192, USA}
\newcommand{\barcelonaa}{Departament de F\'isica Qu\`antica i Astrof\'isica, Universitat de Barcelona, 08028 Barcelona, Spain}
\newcommand{\barcelonab}{Institut de Ci\`encies del Cosmos, Universitat de Barcelona, 08028 Barcelona, Spain}
\newcommand{\ethz}{Institute for Theoretical Physics, ETH Zurich, 8093 Zurich, Switzerland}
\author{Martin~Lebrat}
\thanks{These authors contributed equally to this work}
\author{Muqing~Xu}
\thanks{These authors contributed equally to this work}
\author{Lev~Haldar~Kendrick}
\author{Anant~Kale}
\author{Youqi~Gang}
\affiliation{\harvard}
\author{Pranav~Seetharaman}
\affiliation{\sjsu}
\author{Ivan~Morera}
\affiliation{\barcelonaa}
\affiliation{\barcelonab}
\affiliation{\ethz}
\author{Ehsan~Khatami}
\affiliation{\sjsu}
\author{Eugene~Demler}
\affiliation{\ethz}
\author{Markus~Greiner}
\thanks{\href{mailto:mgreiner@g.harvard.edu}{mgreiner@g.harvard.edu}}
\affiliation{\harvard}

\date{\today}

\begin{abstract}
Quantum interference can deeply alter the nature of many-body phases of matter \cite{auerbach_interacting_2012}. In the paradigmatic case of the Hubbard model, Nagaoka famously proved that introducing a single itinerant charge can transform a paramagnetic insulator into a ferromagnet through path interference \cite{nagaoka_ferromagnetism_1966,thouless_exchange_1965,tasaki_extension_1989}.
However, a microscopic observation of such kinetic magnetism induced by individually imaged dopants has been so far elusive.
Here we demonstrate the emergence of Nagaoka polarons in a Hubbard system realized with strongly interacting fermions in a triangular optical lattice \cite{shastry_instability_1990,white_density_2001}.
Using quantum gas microscopy, we reveal these polarons as extended ferromagnetic bubbles around particle dopants arising from the local interplay of coherent dopant motion and spin exchange.
In contrast, kinetic frustration due to the triangular geometry promotes antiferromagnetic polarons around hole dopants, as proposed by Haerter and Shastry \cite{haerter_kinetic_2005}.
Our work augurs the exploration of exotic quantum phases driven by charge motion in strongly correlated systems and over sizes that are challenging for numerical simulation \cite{anderson_resonating_1973, balents_spin_2010, zhou_quantum_2017}.

\end{abstract}

\maketitle

\section*{Introduction}

Ferromagnetism is a quintessentially quantum phenomenon with subtle origins. Conventionally, it arises from ferromagnetic exchange couplings originating from Coulomb interactions between electrons subject to the Pauli exclusion principle \cite{auerbach_interacting_2012}. This mechanism can however dramatically break down in the presence of strong electronic correlations. A prime example is provided by the Hubbard model, a minimal model capturing interactions between itinerant electrons on a lattice, relevant for a broad range of materials including doped high-temperature superconducting cuprates \cite{lee_doping_2006}. In this model, an antiferromagnetic ground state is favored instead for experimentally relevant interactions at a filling of one particle per site.

Surprisingly, ferromagnetism can be recovered in the limit of infinitely strong interactions by adding one particle dopant to this half-filled state. As first shown by Nagaoka and Thouless \cite{nagaoka_ferromagnetism_1966,thouless_exchange_1965,tasaki_extension_1989}, a ferromagnetic ground state arises from minimizing the kinetic energy of the dopant in a broad class of lattice geometries. Intuitively, Nagaoka ferromagnetism can be understood as the result of constructive interference between different paths the dopant may traverse in the presence of a ferromagnetic spin background (Fig.~\ref{fig:fig1}a). In any other background, dopant tunneling may result in distinguishable spin configurations resulting in paramagnetic or antiferromagnetic states being less energetically favorable (Fig.~\ref{fig:fig1}b).

Nagaoka's exact result however relies on hypotheses that are challenging to meet in realistic materials. Its validity at finite interactions and beyond the single-dopant limit has been the focus of extensive theoretical work \cite{doucot_instability_1989,fang_holes_1989,shastry_instability_1990,basile_stability_1990,barbieri_stability_1990,hanisch_ferromagnetism_1995,wurth_ferromagnetism_1996,white_density_2001,park_dynamical_2008,liu_phases_2012,zhu_doped_2022}. Experimentally, evidence for Nagaoka ferromagnetism was shown on a $2 \times 2$ quantum dot plaquette \cite{dehollain_nagaoka_2020}. Recent quantum simulations of the Hubbard model in Moir\'e heterostructures \cite{tang_simulation_2020,ciorciaro_kinetic_2023} and in cold-atom experiments \cite{xu_frustration-_2023} have mutually supported the existence of magnetic phases with kinetic origin. Despite these advances, signatures of Nagaoka ferromagnetism due to individual dopants in an extended system have so far eluded direct observation.

\placefigure[!t]{f1}

In this work, we experimentally demonstrate the emergence of Nagaoka polarons with strongly interacting ultracold fermions in an optical lattice which pristinely realize the Hubbard model (Fig.~\ref{fig:fig1}c).
These polarons appear as bubbles of enhanced ferromagnetic correlations over areas up to about 30 sites around individual particle dopants, which we image through \emph{in situ} measurements of three-point correlation functions. These bubbles are bounded by antiferromagnetic superexchange occurring at finite interactions (Fig.~\ref{fig:fig1}a), and represent a generalization of Nagaoka’s original arguments \cite{white_density_2001}.
Key to our observations is a triangular optical lattice \cite{xu_frustration-_2023,struck_quantum_2011,yamamoto_single-site-resolved_2020,yang_site-resolved_2021,mongkolkiattichai_quantum_2023,trisnadi_design_2022}, where kinetic magnetism is strongly enhanced due to frustration of antiferromagnetic order and the presence of short-length loops \cite{hanisch_ferromagnetism_1995}. As a result, it is expected to give rise to a variety of single- or few-dopant polaronic states \cite{zhang_pairing_2018,van_de_kraats_holes_2022,davydova_itinerant_2023} expected to be observable through spectroscopic \cite{chen_proposal_2022,morera_spectroscopy_2023} or real-space measurements \cite{morera_high-temperature_2023,schlomer_kinetic_2023,samajdar_nagaoka_2023}.

Triangular lattices furthermore break particle-hole symmetry and Nagaoka's theorem does not apply in the case of a single hole dopant. We however observe evidence for kinetic magnetism around hole dopants in the form of antiferromagnetic bubbles, in agreement with a seminal prediction by Haerter and Shastry \cite{haerter_kinetic_2005}. This asymmetry with respect to doping starkly contrasts with magnetic polarons emerging from exchange-mediated interactions between the dopant and its spin environment \cite{brinkman_single-particle_1970,shraiman_two-particle_1988,sachdev_hole_1989,grusdt_parton_2018}, as investigated in previous cold-atom realizations of the two-dimensional square Hubbard model \cite{koepsell_imaging_2019,ji_coupling_2021,koepsell_microscopic_2021}.
Itinerant spin polarons displaying such a particle-hole asymmetry have recently been observed with ultracold fermions in a triangular lattice \cite{prichard_directly_2023}.

We realize here a frustrated Hubbard model by preparing a balanced mixture of ultracold fermionic lithium-6 atoms in the two lowest hyperfine states and adiabatically loading it into a triangular optical lattice \cite{xu_frustration-_2023}. The tunneling energy $t$ and on-site interaction energy $U$ that solely parameterize our Hubbard quantum simulator are tuned by changing the depth of the optical lattice and by controlling the magnetic field close to the broad Feshbach resonance of lithium-6 (Methods~\ref{subsec:preparation}). This allows us to tune the ratio $U/t$ over more than one order of magnitude from the metallic regime, $U/t = \UDSx$, to the strongly interacting regime $U/t = \UDSv$, where atoms form a large Mott insulator over 300 to 400 sites (Fig.~\ref{fig:fig1}c). Full dopant resolution is obtained by dynamically tuning the lattice geometry to a supersampling square lattice prior to fluorescence imaging (Methods~\ref{subsec:imaging}).

\placefigure[!t]{f2}

\section*{Nagaoka polarons}

We first investigate the regime of small particle doping above half-filling, close to the single-dopant limit required by Nagaoka's theorem. At finite interactions $U$, antiferromagnetic correlations resulting from superexchange coupling $J = 4t^2/U$ can obscure the kinetic magnetism locally induced by dopants. A natural way to quantify this local effect is to measure a connected three-point correlator which captures the amount of magnetism added by particle dopants to the spin background:
\begin{equation} \label{eq:cdss}
    C_\text{dss}(\mathbf{r}_0; \mathbf{d}_1, \mathbf{d}_2) \equiv \frac{4}{\mathcal{N}_\text{dss}} \langle \hat{d}_{\mathbf{r}_0} \, \hat{S}^z_{\mathbf{r}_0 + \mathbf{d}_1} \, \hat{S}^z_{\mathbf{r}_0 + \mathbf{d}_2} \rangle_{c}
\end{equation}
where $\hat{d}_{\mathbf{r}}$ is the doublon occupation operator at site $\mathbf{r}$, $\hat{S}^z_{\mathbf{r}}$ is the projection of the local spin operator along the quantization axis, $\mathcal{N}_\text{dss}$ is the marginal probability to measure a doublon at site $\mathbf{r}_0$ and single spins at sites $\mathbf{r}_0 + \mathbf{d}_1$ and $\mathbf{r}_0 + \mathbf{d}_2$ (see Methods~\ref{subsec:correlators}).
Here and in the following, we measure $C_\text{dss}(\mathbf{r}_0, \mathbf{d}_1, \mathbf{d}_2)$ for pairs of nearest-neighbor spins, $|\mathbf{d}_2 - \mathbf{d}_1| = 1$, and radially average it over all pairs at the same distance $|\mathbf{d}| = |(\mathbf{d}_1 + \mathbf{d}_2)/2|$ from a particle dopant. It is furthermore averaged over a contiguous area of our experimental sample including all sites $\mathbf{r}_0$ at or above half-filling.

We reveal the magnetism induced by a particle dopant on its surrounding spins in spatial maps of the three-point correlator $C_\text{dss}(|\mathbf{d}|)$ in Fig.~\ref{fig:long_range}a. 
At interaction strength $U/t=\UDSi$ where the system is metallic, spin correlations are positive at the shortest distance from the particle dopant. They display a damped oscillation between positive and negative values at longer distances, reminiscent of Friedel oscillations found in Fermi liquids. Such a behavior is suppressed and finally vanishes as interaction strength increases. Strikingly, we find that correlations further from the dopant also turn significantly positive up to a distance $|\mathbf{d}| = 2.5$ at the strongest interaction $U/t=\UDSv$ (Fig.~\ref{fig:long_range}b). These positive correlations form a ferromagnetic bubble covering an area of about 30 sites.

\placefigure[!t]{f3}

We interpret these bubbles of enhanced ferromagnetic correlations as Nagaoka polarons, resulting from mobile particle dopants locally polarizing the antiferromagnetic spin background. 
Initially discussed in the context of the stability of the long-range ferromagnetic state \cite{shastry_instability_1990}, such polarons have been found in the ground state of the square $t-J$ model in density-matrix renormalization group studies \cite{white_density_2001}. Based on variational arguments \cite{auerbach_interacting_2012,white_density_2001}, the radius of the polaron is predicted to weakly scale with interactions as $R_N \sim (t/J)^{1/4} = \mathcal{O}(1)$, which is consistent with an increase of the number of positive correlators $C_\text{dss}$ with $U/t$.

Due to the confinement potential inherent to our trapping laser beams, our region of interest displays a slow spatial variation of the density $n$ between 0.95 and about 1.2. We expect such inhomogeneities to average out the magnitude of the measured correlations and possibly underestimate their range, which makes it difficult to quantitatively estimate the polaron radius. Numerical simulations furthermore show an overall decrease of the range of the correlations with doping, which is consistent with it being limited by the average distance between dopants (see Methods~\ref{subsec:simulations}, Fig.~\ref{fig:numerical_corrmap}).

As the interaction strength $U/t$ is increased, experimental temperatures $T$ remain smaller than the tunneling energy $t$ (Methods~\ref{subsec:calibration}). They however exceed the superexchange energy $J = 4t^2/U$ that determines the magnetic properties of the Hubbard model at half-filling (Fig.~\ref{fig:long_range}c). The persistence of the observed ferromagnetic correlations at large distances in this temperature regime therefore confirms their kinetic origin.

The positive $C_\text{dss}(|\mathbf{d}| = \sqrt{3}/2)$ on the lattice bonds closest to the particle dopant are robust over a wide experimental range of interactions $U/t$ from the metallic to the Mott insulating regime (Fig.~\ref{fig:long_range}d). This robustness may be attributed to the enhancement of quantum interference within the short length-three cycles that compose the triangular lattice. Neglecting interference over longer paths, the ground state of the Hubbard Hamiltonian on a three-site plaquette with one single particle dopant is a triplet state for any interaction strength $U > 0$ (Methods~\ref{subsubsec:plaquette}). In contrast, the ground state on a square plaquette is antiferromagnetic below a critical $U/t \sim 20$ \cite{yao_myriad_2007}, highlighting the tendency to ferromagnetism in short loops. The behavior of the correlator $C_\text{dss}$ both on the closest and next-closest bonds is furthermore qualitatively reproduced by Numerical Linked-Cluster Expansion (NLCE) simulations (Methods~\ref{subsubsec:nlce}).

\placefigure[!t]{f4}

\section*{Kinetic magnetism}

Triangular geometries not only give rise to geometric frustration of antiferromagnetic Heisenberg order but also to kinetic frustration of dopants, with important consequences on dopant-induced magnetism. 
A basic intuition can be gained by considering interference processes on a triangular plaquette (Fig.~\ref{fig:short_range}a), where a dopant exchanges the position of two neighboring spins upon three consecutive tunneling events. 
In contrast to a particle dopant, a hole dopant effectively has a negative tunneling amplitude $-t$, which leads to destructive interference in a ferromagnetic background. 
However, constructive interference can be recovered if the neighboring spin state is antisymmetric upon exchange.
From a kinetic energy perspective, hole dopants therefore favor singlet states with antiferromagnetic correlations.
This asymmetry between particle and hole doping even holds in larger triangular systems: in the infinite $U/t$ limit, Haerter and Shastry predicted the existence of $120^\circ$ antiferromagnetic order around single holes in the ground state \cite{haerter_kinetic_2005} that classically saturates the local magnetic moments \cite{sposetti_classical_2014}, contrary to Heisenberg $120^\circ$ antiferromagnetism driven by superexchange.

Experimentally, we observe antiferromagnetic polarons around single holes as revealed by the hole-spin-spin correlator $C_\text{hss}$, defined similarly to $C_\text{dss}$ in Eq.~\ref{eq:cdss} and plotted in Fig.~\ref{fig:short_range}b, for interaction strength $U/t=\UDSvi$ and densities $0.95 < n < 1.05$. The negative shortest-distance correlations are consistent with the results of Haerter and Shastry, and recent studies \cite{morera_high-temperature_2023, schlomer_kinetic_2023}.
We note that our imaging procedure prevents the measurement of $C_\text{hss}$ between nearest-neighbor spins at larger distances from the hole analogous to Fig.~\ref{fig:long_range}a (Methods~\ref{subsec:correlators}).

After demonstrating the existence of kinetic magnetism carried by polarons close to half-filling, we now explore its evolution as the doping $\delta = n - 1$ is increased. We focus on correlations at the shortest distance $|\textbf{d}| = \sqrt{3}/2$, and first show the total, connected dopant-spin-spin correlators $C_\text{dss}^{\text{tot}}$ and $C_\text{hss}^{\text{tot}}$ in Fig.~\ref{fig:short_range}c, equal to the correlators $C_\text{dss}$ and $C_\text{hss}$ without the uncorrelated normalization factor $\mathcal{N}_\text{dss}$.
At interactions $U/t = \UDSvii$, the non-normalized correlators show a linear doping dependence indicative of a regime where the magnetism induced by each dopant is additive. In this regime, the visible antisymmetry close to half-filling $\delta=0$ between $C_\text{dss}^{\text{tot}}$ for $\delta > 0$ and $C_\text{dss}^{\text{tot}}$ for $\delta < 0$ is due to the Mott insulating nature of the parent system.
Further away from half-filling, the non-normalized correlators decrease in magnitude due to the suppression of the local moments at large dopings but remain positive for doublon dopants and negative for hole-dopants. 
We find good agreement between the experimental data and Determinant Quantum Monte Carlo (DQMC) simulations (Methods~\ref{subsubsec:dqmc}). 
We also show numerical simulations for the non-interacting and $U/t=\infty$ case (Methods~\ref{subsec:simulations}), highlighting the effect of interactions close to half-filling, and the emergence of a linear regime at strong interactions.

Away from half-filling, the asymmetric particle- and hole-induced magnetism is robust to interaction strength, as shown in Fig.~\ref{fig:short_range}d with connected correlators normalized by $\mathcal{N}_\text{dss}$ or $\mathcal{N}_\text{hss}$.
We observe consistently negative short-range correlations $C_\text{hss}$ around holes at all negative dopings $\delta < 0$, and positive short-range correlations $C_\text{dss}$ around particle dopants up to $\delta = +0.5$.
At dopings $|\delta| > 0.2$, non-interacting calculations at a temperature $T/t = 0.5$ display magnitudes similar to experimental data. Numerical simulations of $C_\text{dss}$ from all methods also show quantitative agreement with each other at large particle doping (Fig.~\ref{fig:num_3point}), suggesting that the interaction-dependence of the dopant-spin-spin correlations is weakest in the highly-doped regime.

\section*{Ferromagnetic Transition at~Finite~Doping}

The existence of Nagaoka polarons raises questions about their role in a possible ferromagnetic phase transition when dopant density is increased and polarons start to overlap (Fig~\ref{fig:spin}a). An analogous mechanism has been pointed out theoretically in disordered magnetic semiconductors, where a ferromagnetic transition occurs through the percolation of localized ferromagnetic bubbles as temperature is decreased \cite{kaminski_polaron_2002}.

In our experiment, the sign of the two-point spin correlation function between nearest-neighbors is suggestive of the ferromagnetic or antiferromagnetic nature of the system at equilibrium.
In Fig.~\ref{fig:spin}b we plot the normalized two-point correlator
\begin{equation} \label{eq:css}
    C_\text{ss}(\mathbf{r}; \mathbf{d}) \equiv \frac{4}{\mathcal{N}_\text{ss}} \langle \hat{S}^z_{\mathbf{r}} \, \hat{S}^z_{\mathbf{r} + \mathbf{d}} \rangle_{c}
\end{equation}
$C_\text{ss}$. This correlator is measured as a function of doping at several interaction strengths in the temperature range $T/t=\TDSxiii-\TDSxi$ (Methods, Table~\ref{tab:mega_dataset}).
At half-filling ($\delta = 0$), superexchange interactions lead to an antiferromagnetic state ($C_\text{ss}<0$).
Upon particle doping ($\delta>0$), however, this negative correlation is rapidly suppressed up to a critical doping $\delta_{\text{FM}}$ where it turns positive, consistent with a scenario in which the proliferation of Nagaoka polarons drives a ferromagnetic transition.
Conversely, upon hole doping ($\delta<0$), $C_\text{ss}$ becomes even more negative than at half-filling, consistent with Haerter-Shastry polarons enhancing antiferromagnetism relative to the local moment~\cite{sposetti_classical_2014}. Neither of these trends is present in the equivalent correlator in the square lattice at comparable or larger interaction strengths, plotted in Fig~\ref{fig:spin}c. The latter quantity depends only weakly on doping, consistent with magnetism controlled mainly by the density of moments (that is, $\langle S^z S^z \rangle \propto (1-|\delta|)^2$), in contrast to the kinetic magnetism evident in the triangular lattice.

Decreasing the superexchange energy $J = 4 t^2/U$ by increasing $U/t$ from $\UDSxi$ to $\UDSxiii$ at similar temperatures suppresses superexchange magnetism while preserving kinetic magnetism.
This effect is visible as an upward shift of the $C_\text{ss}$ curve, while its slope versus doping stays roughly constant.
The extreme limit of this effect is captured in Finite-Temperature Lanczos Method (FTLM) simulations at $U/t=\infty$, $T/t=0.6$, where $\delta_\text{FM}=0$.
The net result of this behavior is a rapid reduction of the critical doping $\delta_\text{FM}$ towards half-filling (Fig.~\ref{fig:spin}d) as interactions are increased.
This trend is reminiscent of the existence of a ferromagnetic ground state for an infinitesimal positive doping of one hole in the Nagaoka limit $U/t \rightarrow \infty$, although the experimentally fitted $\delta_\text{FM}$ asymptotically reaches a small finite value, possibly as a consequence of the larger lattice depth and potential gradients due to the trap curvature realized at the strongest interactions.
Numerical simulations at fixed temperature $T/t=0.6$ from the NLCE ($U/t=5$ to $100$, Methods~\ref{subsubsec:nlce}) and FTLM ($U/t=8$ to $\infty$, Methods~\ref{subsubsec:ftlm}) recover qualitatively similar $\delta_{\text{FM}}$ as $U/t$ is increased, but with an asymptote approaching zero as $U/t\to\infty$.

\section*{Discussion and Outlook}
In this work, the enhancement of ferromagnetic correlations with interaction $U$ around single particle dopants (Fig.~\ref{fig:long_range}a, b) and between nearest-neighbor spins (Fig.~\ref{fig:spin}d) suggests that our finite-temperature system forms a precursor to a Nagaoka state at small, positive doping. The robust sign of both dopant-spin-spin (Fig.~\ref{fig:short_range}c, d) and spin-spin correlations (Fig.~\ref{fig:spin}b) away from half-filling furthermore highlights the central role of coherent dopant motion in triangular geometries (Fig.~\ref{fig:fig1}a, b and Fig.~\ref{fig:short_range}a) to stabilize a ferromagnetic state at large particle doping and an antiferromagnetic state at large hole doping, in a regime where spin-exchange magnetism is weak (Fig.~\ref{fig:long_range}d). 

In the infinite-$U$ limit, long-range ferromagnetic order was previously shown to persist in the ground state of the triangular lattice up to remarkably large positive dopings \cite{shastry_instability_1990,hanisch_ferromagnetism_1995} compared to the square one \cite{wurth_ferromagnetism_1996,liu_phases_2012}. At finite interactions, the parent Mott insulating state at half-filling and zero temperature has been conjectured to show a transition from a $120^\circ$ N\'eel ordered state to a quantum spin liquid below $U/t \approx 9-10$, followed by a insulator to metal transition upon further decreasing $U/t$ \cite{szasz_chiral_2020}. Doping the Mott insulator is expected to give rise to competing quantum phases, including chiral metals, spin density waves and superconducting states \cite{weber_magnetism_2006,song_doping_2021,zhu_doped_2022}.

Experimentally, thermal fluctuations associated with our temperatures $T/t \geq 0.3$ prevent long-range order and all measured observables are smooth functions of the interaction strength. Our lowest interactions $U/t \sim 5$ display ferromagnetic nearest-neighbor spin correlations at critical dopings $\delta_\text{FM} > 0.3$ much larger than the strong interaction regime $U/t > 20$. In this weaker interaction regime, finite-doping ferromagnetism might be influenced by a Stoner instability. This hypothesis is supported by numerical Density Matrix Renormalization Group (DMRG) simulations displayed in Fig.~\ref{fig:gs_sim} at $U/t = 20$ showing the formation of a long-range ferromagnetic ground state around a doping of 50\%, where the triangular lattice shows a van Hove singularity. Ferromagnetism in a triangular lattice could show a smooth crossover between the Stoner, mean-field regime at $U/t \approx 0$ and the Nagaoka, $U/t = \infty$, regime. Additional theoretical and numerical studies as a function of lattice geometry and interaction can shed light on this crossover \cite{morera_itinerant_2024}.

Future work can further probe the existence of bound states mediated by kinetic frustration at finite polarization \cite{zhang_pairing_2018,morera_attraction_2021,davydova_itinerant_2023,foutty_tunable_2023} through measurements of spin susceptibility. 
These states have drawn interest from condensed-matter experiments with transition metal dichalcogenides, where observations of kinetic magnetism and spin polarons have recently been reported \cite{tang_simulation_2020,ciorciaro_kinetic_2023,tao_observation_2023}.
Our quantum simulator using ultracold atoms may help elucidate the mechanism of the kinetic magnetism by providing a pristine realization of the triangular lattice Hubbard model and precisely tunable interactions. In the large doping regime, our platform could furthermore investigate dopant pairing and superconductivity based on a `spin-bag' mechanism \cite{schrieffer_spin-bag_1988}. Our lowest experimental temperatures $T/t \approx 0.3$, would allow the exploration of such phenomena governed by tunneling energy $t$, at interactions and dopings where our finite-temperature simulations are challenging over large system sizes.

At weaker interactions where spin exchange becomes dominant, multi-point correlation measurements of spin and density could furthermore help reveal resonating-bond-solid states \cite{majumdar_nextnearestneighbor_2003,auerbach_interacting_2012}. Further decreasing temperature may ultimately elucidate the nature of quantum spin liquid states and intriguing doped phases driven by frustration.

\subsection*{Acknowledgments}
We thank Waseem Bakr, Tilman Esslinger, B.~Sriram Shastry, Richard T. Scalettar, Annabelle Bohrdt, Fabian Grusdt, Henning Schl\"omer and Rhine Samajdar for insightful discussions. We acknowledge support from the Gordon and Betty Moore Foundation, Grant No.~GBMF-11521; National Science Foundation (NSF) Grants Nos.~PHY-1734011, OAC-1934598 and OAC-2118310; ONR Grant No.~N00014-18-1-2863; DOE contract No.~DE-AC02-05CH11231; QuEra grant No.~A44440; ARO/AFOSR/ONR DURIP Grants Nos.~W911NF-20-1-0104 and W911NF-20-1-0163. M.L. acknowledges support from the Swiss National Science Foundation (SNSF) and the Max Planck/Harvard Research Center for Quantum Optics. L.H.K. and A.K. acknowledge support from the NSF Graduate Research Fellowship Program. Y.G. acknowledges support from the AWS Generation Q Fund at the Harvard Quantum Initiative. I.M. acknowledges support from Grant No.~PID2020-114626GB-I00 from the MICIN/AEI/10.13039/501100011033, Secretaria d’Universitats i Recerca del Departament d’Empresa i Coneixement de la Generalitat de Catalunya, cofunded by the European Union Regional Development Fund within the ERDF Operational Program of Catalunya (Project No. QuantumCat, Ref. 001-P-001644). E.K. and P.S. acknowledge support from the NSF under Grant No.~DMR-1918572. E.D. and I.M. acknowledge support from the SNSF project 200021\_212899. 
NLCE calculations were done on the Spartan high-performance computing facility at San Jos\'e State University.

\subsection*{Author Contributions}
M.L., M.X., L.H.K., A.K. and Y.G. performed the experiment and analyzed the data. The numerical simulations were performed by M.X. (DQMC), L.H.K. (noninteracting), A.K. (FTLM), P.S. (NLCE), I.M. (DMRG) and E.K. (NLCE). I.M., E.K. and E.D. developed the theoretical framework. M.G. supervised the study. All authors contributed to the interpretation of the results and production of the manuscript.

\subsection*{Competing Interests}
M.G. is co-founder and shareholder of QuEra Computing.

\begin{dfigure*}{f1}
    \centering
    \noindent
    \includegraphics[width=\linewidth]{"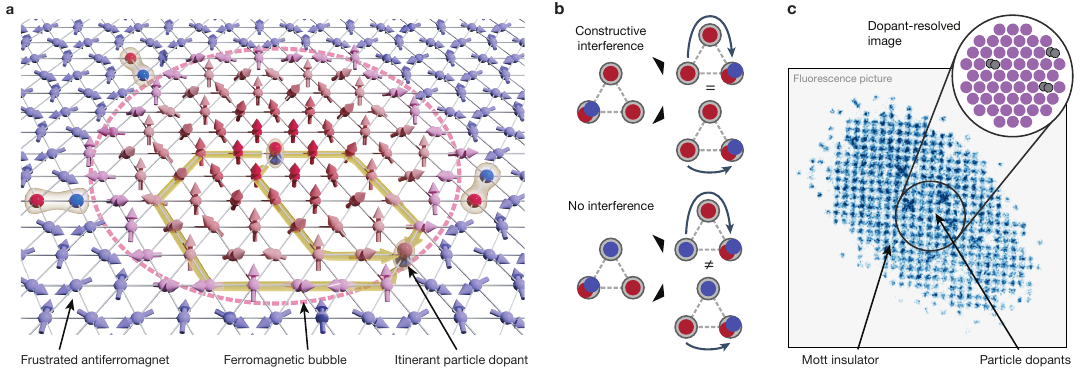"}
    \caption{\textbf{Nagaoka polarons in a Fermi-Hubbard quantum simulator.}
    (\textbf{a})
    As it moves through a half-filled, Mott insulator described by the Hubbard model (purple spins), an itinerant particle dopant (doubly occupied site at center) favors spin alignment in its vicinity (pink circle), giving rise to a ferromagnetic polaron. Ferromagnetism of kinetic origin competes with superexchange coupling (red and blue pairs), which leads to antiferromagnetism away from the dopant.
    In the limit of infinite Hubbard interactions, the radius of the ferromagnetic polaron is expected to diverge and the ground state is a long-range ferromagnet, as shown by Nagaoka. (\textbf{b}) Intuition behind kinetic magnetism: a ferromagnetic background enables constructive interference between paths taken by the dopant, thereby lowering its kinetic energy. Conversely, different paths in a non-polarized background may lead to different spin configurations and reduce quantum interference. (\textbf{c}) We realize the Hubbard model with tunable interaction and tunneling by loading up to about 400 fermionic $^6$Li atoms in a triangular optical lattice. Geometric frustration introduced by the triangular geometry facilitates kinetic ferromagnetism by suppressing superexchange coupling. Particle dopants are imaged at the single-site level (inset) after adiabatically changing the lattice geometry to a supersampling square lattice (shown in the experimental fluorescence picture; see Methods \ref{subsec:imaging}).}
    \label{fig:fig1}
\end{dfigure*}

\begin{dfigure*}{f2}
    \centering
    \noindent
    \includegraphics[width=\linewidth]{"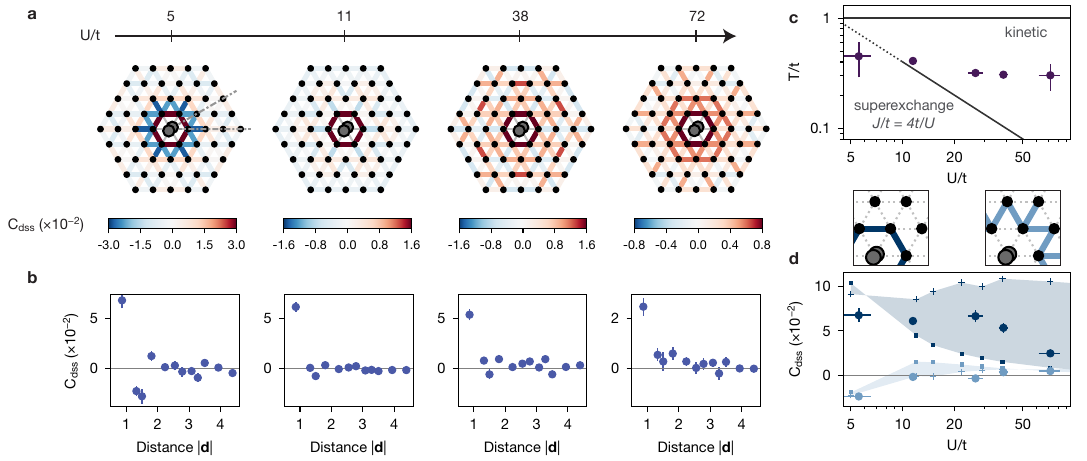"}
    \caption{\textbf{Emergence of Nagaoka polarons around particle dopants.}
    (\textbf{a})
    Quantum gas microscopy enables us to directly investigate how a single dopant affects its spin environment. We observe the resulting ferromagnetic Nagaoka polaron around single doublon dopants by extracting the connected doublon-spin-spin correlation function $C_\text{dss}(|\mathbf{d}|)$ from quantum snapshots.
    The radius of the ferromagnetic bubble increases upon increasing the interaction strength from $U/t = \UDSi$ to $\UDSv$. The three-point correlations are averaged over lattice sites with filling larger than 0.95, then bond averaged according to the spatial symmetries of the triangular lattice, which contains the two reflections indicated with dashed-dotted lines in the leftmost panel. (\textbf{b}) The correlations $C_\text{dss}$ are significantly positive up to a distance $|\mathbf{d}| = 2.5$ from the doubly occupied site at $U/t = \UDSv$. Bonds within a radial distance of 0.15 are averaged together. Here and in the following, errorbars indicate the $1\sigma$ confidence interval. 
    (\textbf{c}) Increasing $U$ suppresses the superexchange energy $J = 4t^2/U$ that determines the magnetic properties of the Hubbard model at half-filling. Our experimental temperatures $T$ in the Mott insulating regime exceed this energy scale while remaining smaller than the tunneling energy $t$, which strongly points at the kinetic origin of the observed magnetic correlations.
    (\textbf{d}) Strong quantum interference on triangular plaquettes results in positive correlations $C_\text{dss}$ at the shortest distance $|\mathbf{d}| = \sqrt{3}/2$ over a wide range of interactions $U/t$ (dark-blue circles; see Methods~\ref{subsubsec:plaquette}), while the sign of the correlator averaged on the next-nearest bonds $|\mathbf{d}| = \sqrt{7}/2$ and $|\mathbf{d}| = 3/2$ reverses between the Fermi liquid and the Mott insulating regimes (light-blue circles). These experimental data at short-range are captured by numerical simulations performed at half-filling (squares) and 3\% particle doping (crosses), performed at temperature $T/t = 0.5$ with Determinant Quantum Monte Carlo (DQMC) for $U/t \leq 12$ and Numerical Linked-Cluster Expansion (NLCE) for $U/t \geq 15$.
    }
    \label{fig:long_range}
\end{dfigure*}

\begin{dfigure*}{f3}
    \centering
    \noindent
    \includegraphics[width=\linewidth]{"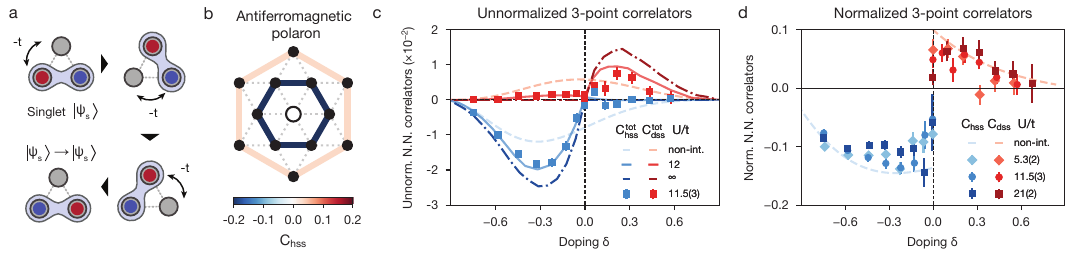"}
    \caption{\textbf{Antiferromagnetic polarons around holes and doping dependence of correlations.}
     (\textbf{a}) In contrast to particle dopants, hole dopants on a triangular plaquette favor antiferromagnetic correlations. Since a hole dopant effectively has a negative tunneling amplitude, it is kinetically frustrated on a triangular lattice with two aligned spins. However, it can delocalize via constructive interference if its neighboring particles are in an antisymmetric spin state upon exchange, \emph{i.e.} in a singlet state. (\textbf{b}) Antiferromagnetic correlations are observed around hole dopants $C_\text{hss}(\mathbf{d}_1, \mathbf{d}_2)$, here averaged in a Mott insulator at interaction strength $U/t=\UDSvi$ and temperature $T/t=\TDSvi$.
     (\textbf{c}) In the Mott insulating regime $U/t=\UDSvi$, the symmetric nature of the ferromagnetic and antiferromagnetic polarons close to half-filling is visible as the linear doping dependence of the unnormalized connected hole (doublon)-spin-spin correlator $C^{\text{tot}}_{\text{hss} ( \text{dss})}$ at dopings $\delta < 0$ ($\delta >0$) between nearest neighbors ($|\mathbf{d}| = \sqrt{3}/2$).
     Away from half-filling, $C^{\text{tot}}_{\text{hss}, \text{dss}}$ decrease in magnitude due to the decreasing local moments.
     The experimental data quantitatively agrees with numerical simulations at $U/t=12$ and $T/t=0.5$.
     (\textbf{d}) Kinetic magnetism at short distances from dopants is robust to doping and varying interaction strength upon normalizing by the uncorrelated part of the correlator, as seen in the negative (positive) nearest-neighbor correlators $C_\text{hss}$ ($C_\text{dss}$). The value of the three-point correlators matches qualitatively the non-interacting calculation away from half-filling.
     The unnormalized correlators for all three interaction strengths is shown in Fig.~\ref{fig:short_range_suppl}. 
     The $U/t = 0, 12,$ and $\infty$ numerics are computed at $T/t = 0.5$ using Wick's contractions, DQMC and Finite-Temperature Lanczos Method (FTLM) respectively (see Methods \ref{subsec:simulations}).
    }
    \label{fig:short_range}
\end{dfigure*}

\begin{dfigure*}{f4}
    \centering
    \noindent
    \includegraphics[width=\linewidth]{"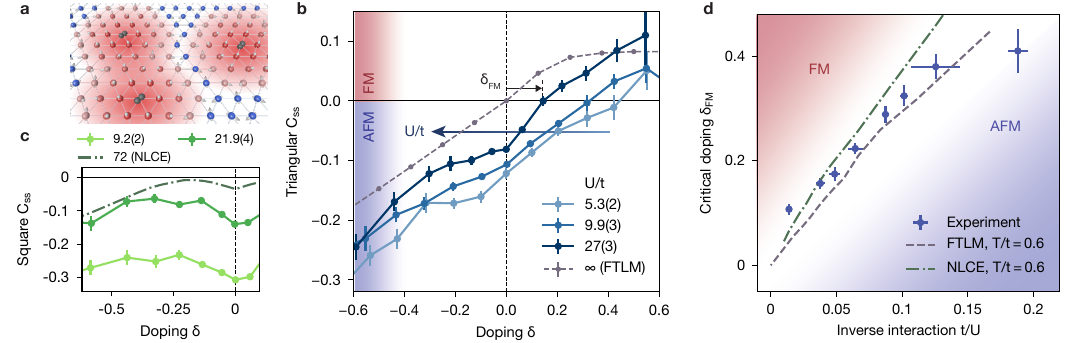"}
    \caption{\textbf{Critical doping for ferromagnetic correlations.}
    (\textbf{a}) Ferromagnetic tendencies increase with particle doping as ferromagnetic polarons start to overlap. (\textbf{b}) A transition from a antiferromagnetic to a ferromagnetic short-range spin background is visible as a change of sign of the normalized nearest-neighbor spin-spin correlations $C_\text{ss}$ in the triangular lattice at a critical doping $\delta_\text{FM}$.
    Numerical simulations at $U=\infty$ and $T/t=0.6$ via FTLM (see Methods~\ref{subsubsec:ftlm}) exhibit a transition at half-filling consistent with Nagaoka's and Haerter-Shastry's pictures, where magnetism is driven by a single dopant.
    (\textbf{c}) No such transition occurs in experimental correlations from the square lattice at comparable $U/t$,
    nor even at $U/t=72$ in correlations computed via NLCE at $T/t=0.4$.
    (\textbf{d}) The critical doping $\delta_\text{FM}$ decreases with increasing $U/t$, both in experimental data (blue), and in numerical simulations via FTLM (dashed line) and NLCE (dash-dotted line). This is consistent with the Nagaoka prediction of a ferromagnetic ground state for an infinitesimal positive doping in the infinite limit $U/t \rightarrow \infty$. Red (blue) shading indicates regions of FM (AFM) correlations in the doping-$t/U$ phase diagram.
    }
    \label{fig:spin}
\end{dfigure*}

\newpage

\setcounter{figure}{0}

\renewcommand{\figurename}{Extended Data Figure}
\renewcommand{\thefigure}{S\arabic{figure}} 
\renewcommand{\tablename}{Extended Data Table}
\renewcommand{\thetable}{\arabic{table}} 

\clearpage

\section*{Methods}

\subsection{Sample preparation}\label{subsec:preparation}
As in previous work~\cite{xu_frustration-_2023}, we prepare an ultracold, spin-balanced gas of $^6$Li in the lowest two hyperfine states and load it into a triangular optical lattice formed by two interfering, actively phase-stabilized beams whose intensities are independently controlled. We refer to these beams as $X$ and $Y$.
We tune the $s$-wave scattering length $a_{s}$ of the Lithium atoms by controlling the magnetic field in the vicinity of the broad Feshbach resonance at $832$G. Combining this and varying the final depth of the lattice allows access to a wide range of $U/t$ values (see Section \hyperref[subsec:calibration]{\textit{Calibration of $t$, $U$, and $T$}}).
In Table~\ref{tab:mega_dataset}, we report the lattice depth and tunneling rate associated with each dataset.
The lattices $1\ldots 6$ are triangular lattices of varying depth, and $7$ is a square lattice.
To ensure the loading remains adiabatic, we use a ramp duration for each final lattice depth that is inversely proportional to the tunneling rate $t$ at the end of the loading ramp. We verify the adiabaticity of the lattice ramp by varying the ramp duration and checking the convergence of the system's density profile as a function of distance from the trap center.
A digital micromirror device (DMD) is used to partially compensate the harmonic confinement created by the Gaussian profile of the lattice beams. In most datasets a parabolic potential is projected from the DMD, although in a few datasets a hyperbolic pattern is projected to further compensate the potential. The potential is thus approximately harmonic in all datasets (see Section \hyperref[subsec:trap-uniformity]{\textit{Trap uniformity and compensation}}).

The experimental datasets thus produced are enumerated in Table~\ref{tab:mega_dataset}, together with the loading parameters, number of shots, interaction strength, and temperature of each dataset and the figures each dataset appears in. The determination of the interaction strength and temperature is described in Section~\hyperref[subsec:calibration]{\textit{Calibration of $t$, $U$, and $T$}}.

\placefigure[!ht]{SA}

\subsection{Imaging procedure and fidelities}\label{subsec:imaging}
To perform measurements on the system after loading it into the lattice, we first freeze the dynamics by quenching the lattice powers in $0.1\rm{ms}$ to 
$(V_X/E_R,V_Y/E_R)\approx(3.2,40)$ (in the notation of~\cite{xu_frustration-_2023}, where $E_R$ is the lattice recoil energy)
at which tunneling is negligible. As in prior work \cite{xu_frustration-_2023}, site-resolved fluorescence imaging is eventually performed on this frozen system in a separate, dedicated imaging lattice.
In the present work, however, we take additional steps before the transfer to the imaging lattice to avoid the issue of parity projection, in which doubly-occupied sites appear empty due to light-assisted collisions during fluorescence imaging \cite{bakr_qgm_2009}.

We achieve this by transferring the atoms from the triangular lattice to a square lattice with twice the number of sites, which converts doubly-occupied sites in the triangular lattice into adjacent pairs of singly-occupied sites in the square lattice.
The transfer is performed by adiabatically ramping up an additional beam, which we call $\bar{X}$ (Fig.~\ref{fig:rampfig}), to about $48E_R$ within $5\rm{ms}$
and ramping off the $X$ lattice at a magnetic field where the interaction between atoms is repulsive. $\bar{X}$ copropagates with $X$, but is detuned in frequency from $X$ and $Y$ by $\sim 1.7\rm{GHz}$. Due to this large frequency offset, $\bar{X}$ effectively does not interfere with $X$ and $Y$, so that the handoff from $X+Y$ to $\bar{X}+Y$ doubles the number of sites ($\bar{X}+Y$ forms a `standard' square lattice). The specific frequency $\sim 1.7\rm{GHz}$ is chosen to position the potential minima of $\bar{X}+Y$ symmetrically relative to those of $X+Y$ in each unit cell of the intermediate $X+\bar{X}+Y$ lattice. This choice minimizes differential potential offsets between the minima of $\bar{X}+Y$ during the handoff, which is necessary to ensure the adiabatic splitting of doubly-occupied sites is robust. The choice of $1.7\rm{GHz}$ is then dictated by wavelength of the lattice light ($1064\rm{nm}$) and the distance to the retroreflection mirror ($\sim4.2\rm{cm}$).

We obtain spin-resolved imaging by selectively removing atoms in one spin state with a resonant laser pulse, as in prior work \cite{parsons_site-resolved_2016}.
In the experiment we use the two lowest hyperfine states of $^{6}$Li, namely the $|F=1/2,m_{F}=1/2\rangle$ and $|F=1/2,m_{F}=-1/2\rangle$ as the effective spin states $|\uparrow\rangle$ and $|\downarrow\rangle$. To prevent doubly-occupied sites from being affected by this pulse, before the splitting procedure we perform a radiofrequency Landau-Zener (LZ) sweep to selectively transfer atoms in $|\downarrow\rangle$ on singly-occupied sites into the $|F=3/2,m_{F}=-3/2\rangle$ ($|3\rangle$) state of the electronic state.
To remove the $|\uparrow\rangle$ state we perform one additional LZ sweep before transferring that exchanges $|\uparrow\rangle$ and $|\downarrow\rangle$ states on singly-occupied sites before transferring $|\downarrow\rangle$ to $|3\rangle$. 

Doubly-occupied sites are not affected by these sweeps due to the interaction-induced shift of the hyperfine transition, which is typically $\sim30 \rm{kHz}$ and hence much larger than the Rabi frequencies of the two sweeps ($390 \rm{Hz}$ and $180 \rm{Hz}$ for the first and second sweeps, respectively).
To ensure adiabaticity we linearly sweep the frequency of the rf signal over a $15 \rm{kHz}$ range centered on the resonance over a duration of $15\rm{ms}$ ($50 \rm{ms}$) for the first (second) sweep.
This third spin state is then targeted for removal using a $10 \mu$s resonant pulse \cite{parsons_site-resolved_2016} after the transfer to the imaging lattice (see Fig.~\ref{fig:rampfig}).

We calibrate the fidelity of fluorescence imaging of the singly-occupied sites $F_{s}=99\%$ as in previous work \cite{xu_frustration-_2023}. To calibrate the fidelity $F_{d,NR}$ of doublons without the LZ transfer and spin removal, we load a cloud of atom with a filling of $n=2$ band insulating state in the center of $\approx 200$ sites.
We find the doublon detection fidelity to be
$F_{d,NR}=98\%$
after reconstruction. To characterize the doublon detection fidelity $F_{d,R\uparrow}$ ($F_{d,R\downarrow}$) in the images with spin $\uparrow$ ($\downarrow$) removed, we apply the same LZ transfer and spin removal pulses used in data taking to the calibration sample with a band insulator core and find the fidelity to be
$F_{d,R\uparrow(R\downarrow)}=95\%$.

\placefigure[!t]{SB}
\placefigure[!tp]{SC}

\subsection{Calibration of $t$, $U$, and $T$}\label{subsec:calibration}
We obtain $U/t$ and $T/t$ in the triangular lattice by comparing experimental double occupancy densities and spin correlations to determinant quantum Monte Carlo (DQMC) and finite temperature Lanczos method (FTLM) simulations.
For datasets from the square lattice, we use a similar procedure but compare against numerical linked cluster expansion (NLCE) simulations from Ref.~\cite{nlce3}.
The results for $U/t$ and $T/t$ are listed in Table~\ref{tab:mega_dataset}
.

As described in Section \hyperref[subsubsec:dqmc]{\textit{Determinant Quantum Monte Carlo (DQMC)
simulation}}, we perform DQMC simulations of the triangular lattice Hubbard model on a mesh of $\mu$, $U$, and $T$, with simulation parameters described. At each point in the mesh we compute the particle density $n(\mu,U,T)$, the double occupancy $d(\mu,U,T)$, and the nearest-neighbor spin correlator $C_\text{ss}(\mu,U,T)$. As $U/t$ increases, DQMC becomes less stable due to the sign problem. However, for $U/t > 20$, we found the sign problem is absent at half-filling $n=1$ and can be computed down to a temperature of $T/t=0.3$. 

To obtain $U/t$, we first perform linear interpolation on DQMC data using the experimentally measured double occupancy $d$ and the nearest-neighbor spin correlator $C_\text{ss}$ at half-filling. Since we have a spatially varying atom density (see Section \hyperref[subsec:trap-uniformity]{\textit{Trap uniformity and compensation}}), these half-filling observables are determined by averaging over lattice sites with average density within $[0.97, 1.03]$, which is the most narrow range that includes enough lattice sites to reduce statistical noise. However, the values of $U/t$ from interpolation still vary between datasets with the same lattice parameters due to statistical noise, and at $U/t > 20$, double occupancy $d$ decreases to $<1\%$ and is more susceptible to imaging infidelity. Thus, we correct the interpolated value using the linear dependence of $U$ on scattering length $a_{s}$~\cite{bloch_dalibard_zwerger_RMP}.

For datasets with final calibrated $U/t < 20$, we take several other datasets with the same lattice depth and different magnetic field, and we perform a linear fit of interpolated $U/t$ on $a_{s}$, using values for $a_s$ from Ref.~\cite{Zurn_scattering_length}. For datasets with final calibrated $U/t > 20$, we take data at the same lattice depth but smaller magnetic field where $U/t$ is small and the measured double occupancy is still a faithful parameter to calibrate $U/t$, and then scale the interpolated $U/t$ proportionally by $a_{s}$. This method produces the same $U/t$ for the same lattice parameters and is robust against imaging infidelity.

To obtain $T/t$, we perform similar linear interpolation on DQMC data, but using the calibrated $U/t$ and the experimentally measured nearest-neighbor spin correlator $C_\text{ss}$ at half-filling. However, for $U/t > 35$, the sign problem of DQMC becomes severe even at half-filling, so we perform similar interpolation based on FTLM simulation of the $t-J$ model, as described in Section \hyperref[subsubsec:ftlm]{\textit{Finite Temperature Lanczos Method (FTLM) simulation}}.

In datasets from the square lattice, we obtain the experimental $C_\text{ss}$ and $d$ at half-filling as above. We then determine both $U/t$ and $T/t$ by linear interpolation on NLCE data from Ref.~\cite{nlce3}.

We obtain the absolute value of the tunneling $t$ in Hz as described in previous work~\cite{xu_frustration-_2023}. We report the resulting tunneling rates in Table~\ref{tab:mega_dataset}. Note that lattice $7$ is a square lattice, while lattices $1\ldots 6$ are triangular.

\placefigure[!t]{SD}

\subsection{Correlation functions}\label{subsec:correlators}

\subsubsection{Definition}

The normalized, connected doublon-spin-spin correlator used in the main text is defined as:
\begin{equation} \label{eq:normalized-three-point-correlator}
    C_\text{dss}(\mathbf{r}_0; \mathbf{d}_1, \mathbf{d}_2) \equiv \frac{4}{\mathcal{N}_\text{dss}} \langle \hat{d}_{\mathbf{r}_0} \, \hat{S}^z_{\mathbf{r}_0 + \mathbf{d}_1} \, \hat{S}^z_{\mathbf{r}_0 + \mathbf{d}_2} \rangle_c
\end{equation}
Here and in the following, the factor of 4 is used to normalize the $\hat{S}^z \hat{S}^z$ part of the correlator to one. 
The denominator is defined as $\mathcal{N}_\text{dss} = \langle d \rangle \langle p \rangle^2$, where $\langle d \rangle$ ($\langle p \rangle$) is the average probability for a site to be doubly (singly) occupied, and provides an upper bound for the three-point correlator.
$\langle ... \rangle_c$ denotes the connected part of the three-point correlation function, that is the difference between the doublon-spin-spin correlator and its disconnected parts. 
Under the assumption of a spin-balanced atomic mixture with total spin projection along $z$ $\langle S^z \rangle = 0$, this connected correlator simplifies to:
\begin{multline} \label{eq:connected-three-point-correlator}
C_\text{dss}(\mathbf{r}_0; \mathbf{d}_1, \mathbf{d}_2) = \frac{4}{\mathcal{N}_\text{dss}} \langle \hat{d}_{\mathbf{r}_0} \, \hat{S}^z_{\mathbf{r}_0 + \mathbf{d}_1} \, \hat{S}^z_{\mathbf{r}_0 + \mathbf{d}_2} \rangle\\
- C_\text{ss}(\mathbf{r}_0+\mathbf{d}_1, \mathbf{r}_0+\mathbf{d}_2).
\end{multline}
It can be interpreted by the amount of spin correlations added by doublons to the normalized spin correlation background:
\begin{equation}
    C_\text{ss}(\mathbf{r_1}, \mathbf{r_2}) = \frac{4}{\mathcal{N}_\text{ss}} \langle S^z_\mathbf{r_1} S^z_\mathbf{r_2} \rangle
\end{equation}
with normalization factor $\mathcal{N}_\text{ss} = \langle p \rangle^2$. 
Similarly, the normalized, connected hole-spin-spin correlator is defined as
\begin{equation} \label{eq:normalized-hss-correlator}
    C_\text{hss}(\mathbf{r}_0; \mathbf{d}_1, \mathbf{d}_2) \equiv \frac{4}{\mathcal{N}_\text{hss}} \langle \hat{h}_{\mathbf{r}_0} \, \hat{S}^z_{\mathbf{r}_0 + \mathbf{d}_1} \, \hat{S}^z_{\mathbf{r}_0 + \mathbf{d}_2} \rangle_c, 
\end{equation}
where $\mathcal{N}_\text{hss} = \langle h \rangle \langle p \rangle^2$, and $\langle h \rangle$ is the average probability for a site to be empty (hole).

In Fig.~\ref{fig:short_range}d we show the nearest neighbor three-point correlator $C_\text{dss}$ defined as 
\begin{multline} \label{eq:normalized-dss-correlator-3d}
    C_\text{dss}(\delta) \equiv \frac{1}{\mathcal{N}_{\Omega_\delta}}\sum_{\mathbf{r}_0\in \Omega_\delta}\frac{1}{3}\Big(C_\text{dss}(\mathbf{r}_0; \mathbf{e}_1, \mathbf{e}_2) +\\
    C_\text{dss}(\mathbf{r}_0; \mathbf{e}_2, \mathbf{e}_3)
    + C_\text{dss}(\mathbf{r}_0; \mathbf{e}_3, \mathbf{e}_1)\Big),
\end{multline}
where $\Omega_\delta$ is a region of with average doping level $\delta$, $\mathcal{N}_{\Omega_\delta}$ is the number of lattice sites in this region, and $\mathbf{e}_1, \mathbf{e}_2,$ and $\mathbf{e}_3$ are the three unit vectors along the triangular lattice bonds. We similarly define the nearest neighbor hole-spin-spin correlator $C_\text{hss}$.
In Fig.~\ref{fig:short_range}c we also show the non-normalized three-point correlators $C_\text{dss/hss}^{\text{tot}}$ which are defined without the normalization factors ${\mathcal{N}_\text{dss/hss}}$:
\begin{equation} \label{eq:nonnormalized-three-point-correlator}
    C^\text{tot}_\text{dss}(\mathbf{r}_0; \mathbf{d}_1, \mathbf{d}_2) \equiv 4 \langle \hat{d}_{\mathbf{r}_0} \, \hat{S}^z_{\mathbf{r}_0 + \mathbf{d}_1} \, \hat{S}^z_{\mathbf{r}_0 + \mathbf{d}_2} \rangle_c
\end{equation}

In Fig.~\ref{fig:spin} we show the nearest-neighbor spin correlation $C_\text{ss}$ defined as 
\begin{multline}
    C_\text{ss}(\delta) \equiv \frac{1}{\mathcal{N}_{\Omega_\delta}}\sum_{\mathbf{r}_0\in \Omega_\delta}\frac{1}{3}\Big(C_\text{ss}(\mathbf{r}_0, \mathbf{r}_0 +  \mathbf{e}_1)\\
    + C_\text{ss}(\mathbf{r}_0, \mathbf{r}_0 + \mathbf{e}_2)
    + C_\text{ss}(\mathbf{r}_0; \mathbf{r}_0 + \mathbf{e}_3)\Big),
\end{multline}
where we average the correlator over the three equivalent lattice bonds and over a region of constant doping level. 

\subsubsection{Computation from experimental snapshots}\label{subsubsec:experimental_correlators}
As described in the previous section, we experimentally obtain three types of snapshots: (i) with no spin removal ($NR$), (ii) after removing spin $\uparrow$ ($R\uparrow$) and (iii) after removing spin $\downarrow$ ($R\downarrow$). 
Further, in all three sets of images we can distinguish $0$, $1$ and $2$ atoms per site which we label as $h$, $p$ and $d$ respectively.
Table \ref{tab:density_and_spin_removal_summary} lists the site-resolved observed outcomes and the possible site occupations that map to the same measured outcome. 
With these three sets of images (even with parity projected imaging) we can obtain connected two point spin correlators $\avg{S^z_i S^z_j}_c$ for arbitrary sites $i$ and $j$ as demonstrated in our previous work \cite{parsons_site-resolved_2016}. 
We repeat the formula below for clarity:
\begin{equation} \label{eq:spin-spin-correlator-formula}
\begin{split}
    4\avg{S^z_i S^z_j}_c &= 2\sum_{\sigma \in \{\uparrow, \downarrow\}}{\avg{p_i p_j}^{(R\sigma)}_c } - \avg{p_i p_j}^{(NR)}_c,
\end{split}
\end{equation}
where $\avg{~}^{(NR)}$ refers to the expectation value over multiple images where neither spin is removed, and $\avg{~}^{(R\sigma)}$ refers to the expectation value over images where atoms in spin $\sigma$ are removed.

With the addition of full density resolution we can also obtain connected three point correlator doublon-spin-spin $\avg{d_i S^z_j S^z_k}_c$ for arbitrary sites $i,j$ and $k$ using the formula:
\begin{equation} \label{eq:doublon-spin-spin-formula}
\begin{split}
    4\avg{d_i S^z_j S^z_k}_c &= 2\sum_{\sigma \in \{\uparrow, \downarrow\}}{4\avg{d_i p_j p_k}^{(R\sigma)}_c } - 4\avg{d_i p_j p_k}^{(NR)}_c.
\end{split}
\end{equation}
This formula is a simple modification of eq. \ref{eq:spin-spin-correlator-formula} since we can uniquely identify doublons in each of the three sets of images as seen in Table \ref{tab:density_and_spin_removal_summary}. 

On the other hand, since holes cannot be uniquely identified in our imaging scheme (a hole observed in a spin-removal image could be a hole or the spin which was removed), we cannot construct the hole-spin-spin correlator $\avg{h_i S^z_j S^z_k}_c$ for arbitrary sites $i,j$ and $k$. 
However, we can still obtain a permutation symmetrized combination of correlators $C_{\text{hss}}({i,j,k}) = \sum_{(\bar{i},\bar{j},\bar{k})\in (i,j,k)}{\avg{h_{\bar{i}} S^z_{\bar{j}} S^z_{\bar{k}} }_c}$ using the following formula:
\begin{multline} \label{eq:hole-spin-spin-formula}
    C_{\text{hss}}({i,j,k}) = \\
    \frac{4}{9}\sum_{(\bar{i},\bar{j},\bar{k})\in (i,j,k)}\Bigg(
    2\sum_{\sigma \in \{\uparrow, \downarrow\}}{\left(\avg{h_{\bar{i}} p_{\bar{j}} p_{\bar{k}}}^{(R\sigma)}_c - \avg{h_{\bar{i}} h_{\bar{j}} p_{\bar{k}}}^{(R\sigma)}_c \right)} \\
    + \left(\avg{h_{\bar{i}} p_{\bar{j}} p_{\bar{k}}}^{(NR)}_c + 2 \avg{h_{\bar{i}} h_{\bar{j}} p_{\bar{k}}}^{(NR)}_c \right)\Bigg) .
\end{multline}
We can see how this formula works by writing out the first term of the connected correlator in the occupation basis. 
For convenience, we drop the site labels and imply averaging over cyclic permutation of the three sites. 
\begin{equation}\label{eq:hss-LHS}
4\avg{h S^z S^z} = \avg{h \uparrow \uparrow} + \avg{h \downarrow \downarrow} - \avg{h \uparrow \downarrow} - \avg{h \downarrow \uparrow}.
\end{equation}
Similarly writing out the three body terms from Eq.~\ref{eq:hole-spin-spin-formula}, 
\begin{equation} \label{eq:hss-expanded}
\begin{split}
    & \frac{2}{3}\sum_{\sigma}{\left(\avg{h p p}^{(R\sigma)} - \avg{h h p}^{(R\sigma)}\right)}\\
    &~~+ \frac{1}{3}\left(\avg{h p p}^{(NR)} + 2 \avg{h h p}^{(NR)}\right)\\
    &= \frac{2}{3}\Big(
    \avg{\uparrow \downarrow\downarrow} + \avg{h \downarrow\downarrow} + \avg{\downarrow \uparrow\uparrow} + \avg{h \uparrow\uparrow} \Big) \\
    &~~-\frac{2}{3} \Big(\avg{h h \downarrow} + \avg{h \uparrow \downarrow} + \avg{\uparrow h \downarrow} + \avg{\uparrow \uparrow \downarrow} \\
    &~~+ \avg{h h \uparrow} + \avg{h \downarrow \uparrow} + \avg{\downarrow h \uparrow} + \avg{\downarrow \downarrow \uparrow}\Big)\\
    &~~+ \frac{1}{3}\Big(\avg{h \uparrow \uparrow} + \avg{h \uparrow \downarrow} + \avg{h \downarrow \uparrow} + \avg{h \downarrow \downarrow}\Big) \\
    &~~+ \frac{2}{3} \Big(\avg{h h \uparrow} + \avg{h h \downarrow}\Big) \\
    &= \frac{2}{3}\Big(\avg{h \downarrow\downarrow}  + \avg{h \uparrow\uparrow} \Big) \\
    &~~-\frac{2}{3} \Big(+ \avg{h \uparrow \downarrow} + \avg{\uparrow h \downarrow}
    + \avg{h \downarrow \uparrow} + \avg{\downarrow h \uparrow} \Big)\\
    &~~+ \frac{1}{3}\Big(\avg{h \uparrow \uparrow} + \avg{h \uparrow \downarrow} + \avg{h \downarrow \uparrow} + \avg{h \downarrow \downarrow}\Big)\\
    &= \avg{h \uparrow \uparrow} + \avg{h \downarrow \downarrow} - \avg{h \uparrow \downarrow} - \avg{h \downarrow \uparrow},
\end{split}
\end{equation}
where we used the cyclic permutation to cancel terms in the last step. 
Similarly, in the two body and on-site terms appropriate terms get cancelled after cyclic permutation to give eq. \ref{eq:hole-spin-spin-formula}.

\placefigure[!t]{SE}

\subsection{Trap uniformity and compensation}\label{subsec:trap-uniformity}

\subsubsection{Trap curvature}
Due to the Gaussian envelope of the lattice beams and the additional light projected from the DMD, the atoms experience a spatially varying chemical potential.
In a given dataset, we may estimate the resulting potential gradients by measuring the average experimental density $\langle n_{\mathbf{r}}\rangle$ on each site, and using the equation of state computed in DQMC to extract the local chemical potential as $\mu(\mathbf{r})=\mu_\text{DQMC}(\langle n_{\mathbf{r}}\rangle,U,T)$ within the local density approximation. Here $U$ and $T$ are obtained as in Sections \hyperref[subsec:calibration]{\textit{Calibration of $t$, $U$, and $T$}}, and the equation of state is inverted by linear interpolation on DQMC data from a range of $\mu$ values. We quantify the potential gradients by fitting parabolae to cuts of the local potential along the cloud's major and minor axes, $\mu(r)=\mu_0-\frac{1}{2}\kappa r^2$. We report the fitted trap curvatures $\kappa_{\rm{maj}}$ and $\kappa_{\rm{min}}$ in Table~\ref{tab:curvature} for a representative subset of datasets. The uncertainties on these numbers account both for uncertainties on the site-resolved density $\langle n(\mathbf{r})\rangle$ as well as on $U$ and $T$.
Typically $\kappa_{\rm{maj}}$ ranges from $0.03-0.09t/\rm{sites}^2$, while $\kappa_{\rm{min}}$ ranges from $0.3-0.6t/\rm{sites}^2$.

In datasets DS6, DS8, and DS10, where we report (Fig.~\ref{fig:short_range}) three-point correlators at finite hole doping, we project a hyperbolic potential from the DMD to partially compensate the minor axis confinement. The values of $\kappa_{\rm{min}}$ are thus correspondingly lower, for example, in DS10 and DS6 than in DS11 and DS7 (the analogous datasets for doublon doping in Fig.~\ref{fig:short_range}).
This is done to reduce the local gradient in the hole-doped region of the trap, which can alter the value of correlation functions if it is too strong, as is discussed below. Such compensation is unnecessary in the doublon-doped datasets because the doublon-doped region naturally occurs close to the trap center.

\subsubsection{Sensitivity to potential gradients}

To estimate the sensitivity of the correlations reported in this paper to the potential gradients produced by the harmonic confinement, we perform FTLM simulations (see Section \hyperref[subsubsec:ftlm]{\textit{Finite Temperature Lanczos Method simulation}}), which we expect to qualitatively capture the relative effects of a potential gradient.

An example of the results of these calculations is shown in Fig.~\ref{fig:gradient}, obtained from a $4\times 3$ $t-J$ cluster at fixed $U/t=30$ and $T/t=0.5$. The left column plots the nearest-neighbor $C_\text{ss}$ and smallest-triangle $C_{\text{hss},\text{dss}}$ correlators (see Section~\ref{subsec:correlators}) as a function of density at selected values of the gradient $\Delta$ (measured in $t/\rm{site}$).
The right column plots the same quantities as a function of the gradient strength at three example densities, chosen to be below, above, or at half-filling.
In this regime, we find the spin-spin correlations close to a Mott insulator remain robust even in the presence of gradients $\sim10t/\rm{site}$.
The spin-spin correlations at finite doping and three-point correlations, however, are more significantly affected by gradients.

This difference in sensitivity to gradients in doped vs undoped systems reflects the kinetic nature of the magnetism at finite doping. 
Dopant mobility is reduced in the presence of gradients, due to the suppression of resonant tunneling by site-to-site potential offsets.
Potential gradients thus suppress kinetic magnetism, which results from the motion of dopants.
In contrast, the virtual tunneling responsible for superexchange interactions at half-filling in a Mott insulator is relatively unaffected by potential gradients (see for example \cite{hirthe_magnetically_mediated,annabelle_mixD}).
As a result, magnetism at half-filling is much more robust to potential gradients than it is at finite doping.

Qualitatively, the most significant consequence of potential gradients is a strong reduction and even reversal of the hole-spin-spin correlator $C_\text{hss}$ on the hole-doped side, which we address by specifically using trap compensation in the related dataset of Fig.~\ref{fig:short_range}. Quantitatively, we expect gradients to overall decrease the doublon-spin-spin correlators on the particle-doped side (Figs.~\ref{fig:long_range} and \ref{fig:short_range}), to decrease the magnitude of the spin-spin correlator $C_\text{ss}$ on the hole-doped side, and to increase of the critical particle doping $\delta_\text{FM}$ where the spin-spin correlator $C_\text{ss}$ becomes ferromagnetic (Fig.~\ref{fig:spin}).

\subsection{Doping- and interaction-dependence of the doublon-spin-spin correlations}
As described in the main text and in Section~\hyperref[subsec:trap-uniformity]{\textit{Trap uniformity and compensation}}, the region of interest used to compute the doublon-spin-spin correlation maps shown in Fig.~\ref{fig:long_range} displays small spatial variations of the atom density. This leads to an averaging of the correlations $C_\text{dss}$ over different doping values. To evaluate the exact doping dependence of the correlations, we perform numerical simulations at a reference temperature $T/t \approx 0.5$, see Fig.~\ref{fig:numerical_corrmap}. Overall, the magnitude of $C_\text{dss}$ at large distances from the dopant is suppressed for dopings $\delta > 5\%$, which is generally expected to reduce the magnitude and range of the experimental correlations after averaging. 

At small to moderate interactions $U/t = 5$ and $12$, we perform DQMC to obtain all dopant-spin-spin correlators on a $8 \times 8$ system size and for a large range of dopings. At $U/t = 5$, a Friedel-type oscillatory behavior is clearly visible as a function of distance from the doublon dopant or all the doping values shown in Fig.~\ref{fig:numerical_corrmap}a. The magnitude of the correlation is reduced for dopings $\delta > 20\%$. At $U/t=12$, $C_\text{dss}(|\mathbf{d}|)$ for $|\mathbf{d}|>1$ start to be positive close to half-filling but turn negative at dopings $\delta\sim2\%$ (Fig.~\ref{fig:numerical_corrmap}b). Notably, the $2^\text{nd}$ and $3^\text{rd}$ nearest neighbor turns negative at smaller doping than the $4^\text{th}$.

At larger interactions, we turn to NLCE simulations as the sign problem of DQMC becomes severe. We implemented $C_\text{dss}(|\mathbf{d}|)$ up to the $5^{\mathrm{th}}$ nearest-neighbor from the doublon, with all further neighbors are set to be $0$ in the plot. 
At half-filling, all correlations are vanishing with interaction, in strong contrast with the particle-doped case (Fig.~\ref{fig:numerical_corrmap}c-e).

In Fig.~\ref{fig:nlce_fig2}, we highlight the evolution of the $1^\text{st}$ nearest neighbor $C_\text{dss}$ and the average of the $2^\text{nd}$ and $3^\text{rd}$ nearest-neighbor correlations (as defined in Fig.~\ref{fig:long_range}), computed with NLCE across all interactions at doping values $\delta=0.0$ and $\delta=0.05$. At half-filling, the Hubbard model effectively maps to a Heisenberg model in the limit of large $U/t$ and since the simulation is performed at constant temperature $T/t$, the effective increase of the temperature relative to superexchange $T/J$ leads to a decrease of correlations. In contrast, at finite doping, the $1^\text{st}$ nearest-neighbor correlator shows a very weak dependence on interaction strength, and the $2^\text{nd}$ turns from negative to positive at $U/t\sim 30$. This provides another confirmation that magnetism away from half-filling is not governed by superexchange $J$ but by the presence of mobile particle dopants with kinetic energy $t$.

At the largest interactions, doublon-spin-spin correlations also display a non-monotonic behavior as doping is increased: both the  range and the absolute value of the correlations increase at all $5$ distances up to $\delta\sim1\%$ doping and then start to decrease (up $8\%$ where NLCE starts to become unstable at $U/t = 100$), with a notably weak $3^\text{rd}$ nearest-neighbor correlator.

One difference between Fig.~\ref{fig:long_range}d and Fig.~\ref{fig:nlce_fig2} is the more abrupt decrease of the experimentally measured $1^\text{st}$ nearest-neighbor at $U/t = 72(11)$. We attribute it to the experimental increase of the size of the Mott insulator region with interactions, which leads to a stronger weighting of weak correlations close to half-filling when performing a spatially uniform average.

\subsection{Data analysis}\label{subsec:analysis}
In Fig.~\ref{fig:long_range}, three-point correlation functions are computed over all triplets of sites whose average density is above 0.95. This corresponds to a spatial average over 123 sites at $U/t = \UDSi$ in the metallic regime, and over 271 to 330 sites at the four other interactions in the Mott insulating regime. To eliminate slow shot-to-shot variations of the atom number that may introduce systematical shifts in the computed correlations, experimental images are postselected within a window of $\pm 15$ atoms away from the mean atom number for fully density-resolved images, and $\pm 10$ atoms for spin-resolved images (corresponding to relative fluctuations of about $\pm 5\%$ in both cases). 

The data in Figs~\ref{fig:short_range} and ~\ref{fig:spin} are postselected with a window of $\pm 30$ atoms in fully density-resolved images and $\pm 20$ atoms in spin-resolved images.
Experimental curves versus doping are obtained by binning the sites of the lattice according to their measured density and averaging correlation functions within each bin, with typically 50 sites per bin.
The experimental value of $\delta_{\text{FM}}$ in Fig.~\ref{fig:spin}d is obtained from the zero of a linear fit to the curve thus obtained for $C_{\rm{ss}}$ in each dataset.

All errorbars indicate the $1\sigma$ confidence interval obtained by using bootstrap sampling across all experimental snapshots of a given dataset with $100$ randomly generated samples.

\subsection{Comparison between numerical methods}

We show the numerically computed normalized doublon-spin-spin and hole-spin-spin correlators in Fig.~\ref{fig:num_3point}a-d for finite temperature $T/t=1$ as well as in the ground state.
Surprisingly, we find an almost universal behavior of the normalized doublon-spin-spin correlator above half-filling with very weak $U/t$ dependence above doping $\delta>0.1$.
Similarly we find weak $U/t$ dependence of the normalized hole-spin-spin correlator below doping $\delta<-0.1$. 
However, as seen in Fig.~\ref{fig:long_range} and Fig.~\ref{fig:numerical_corrmap}, the three point correlators beyond nearest neighbor vary significantly with interaction strength and show the range of the Nagaoka polaron increasing with interactions.

For completeness, we also show in Fig.~\ref{fig:num_3point} the numerically computed bare doublon-spin-spin and hole-spin-spin correlator for neareast neighbors (panels e and f) and as well as the non-normalized, connected correlators (panels g and h). 
The bare correlator doublon-spin-spin is defined as the first term in Eq.~\ref{eq:connected-three-point-correlator} without subtracting out the disconnected terms (similarly for the bare hole-spin-spin correlator). 
We note that the bare doublon-spin-spin correlator shows a sign change on the going from low $U/t$ to large $U/t$ (similar to the spin-spin correlator). Whereas the connected correlator is positive for all dopings and interaction strengths, indicating that the local spin correlations added by doublons is always ferromagnetic for all $U/t > 0$.

\placefigure[!t]{SF}

\subsection{Doping-induced long-range ferromagnetism}

The experimental observation of the Nagaoka polaron paves the way towards the detection of kinetic induced long-range ferromagnetism. Each doublon induces around it a small ferromagnetic region, forming the Nagaoka polaron. When multiple doublons are injected into the system and the corresponding Nagaoka's polarons start to overlap, a transition towards a long-range ferromagnet occurs. To show the formation of long-range ferromagnetism upon doublon doping we numerically compute the total spin squared at zero temperature,
\begin{align}
    \langle \mathbf{S}^2 \rangle = \sum_{ij} \langle \mathbf{S}_i\mathbf{S}_j \rangle .
\end{align}
Then, we associate a net total spin $\langle \mathbf{S} \rangle$ via the relation $\langle \mathbf{S}^2 \rangle = \langle \mathbf{S} \rangle \left(\langle \mathbf{S} \rangle +1 \right)$.

A long-range SU$(2)$ ferromagnet is characterized by exhibiting a maximum total spin $\langle S \rangle = \left(N_s-N_D \right)/2$, where $N_s$ is the number of sites and $N_D$ is the number of dopants $N_D=|\delta| N_s$. In Fig.~\ref{fig:gs_sim}a we show the dependence of the total spin as a function of the doping for a strong on-site interaction $U/t=20$. We observe a transition towards a long-range ferromagnet at a critical doublon doping $\delta_{c_1}\sim 0.45$. Moreover, the ferromagnetic state becomes unstable at a larger doublon doping $\delta_{c_2}\sim 0.6>\delta_{c_1}$. Our numerical results support a scenario where the overlap of multiple Nagaoka's polarons gives rise to the emergence of a long-range ferromagnetic state in the strongly interacting regime at zero temperature. Its interplay with other mechanisms (such as Stoner or flat-band ferromagnetism) as a function of the interaction $U/t$ and system's geometry is investigated in~\cite{morera_itinerant_2024}.

\subsection{Numerical methods}\label{subsec:simulations}

In the main text and the following, we define the Hubbard Hamiltonian as follows:
\begin{equation*} \label{eq:Hubbard-Hamiltonian}
    \hat{\mathcal{H}} = -t\sum_{\langle i, j \rangle,\sigma} (\hat{c}^\dagger_{i} \hat{c}_{j,\sigma} + \text{h.c.}) + U \sum_{i} \hat{n}_{i,\uparrow} \hat{n}_{i,\downarrow}-\sum_{i,\sigma} \mu_i \hat{n}_{i,\sigma}
\end{equation*}
where $\hat{c}^{(\dagger)}_{i,\sigma}$ denotes the fermionic annihilation (creation) operator for spin $\sigma = \uparrow, \downarrow$ on lattice site $i$. The first sum is performed over pairs of nearest-neighbor sites $\langle i, j \rangle$. We chose the convention $t > 0$, leading to a negative tunneling amplitude for particles and a positive tunneling amplitude for holes.

\subsubsection{Toy model on a triangular plaquette}\label{subsubsec:plaquette}
Insights on the microscopic processes behind kinetic magnetism can be gained by considering the previous Hamiltonian over a triangle formed by three sites $i = 0, 1, 2$.

The case of a single hole dopant on a half-filled plaquette restricts the Hilbert space to Fock states consisting in two single spins or one doublon ($\ket{0}$ is the vacuum state with no particle):
\begin{align*}
\ket{{\sigma \sigma'}_i} &= \hat{c}^\dagger_{i+1, \sigma} \hat{c}^\dagger_{i+2, \sigma'} \ket{0} \\
\ket{D_i} &= \hat{c}^\dagger_{i, \uparrow} \hat{c}^\dagger_{i, \downarrow} \ket{0}.
\end{align*}

Rewriting the two-spin states as triplet and singlet eigenstates of the total spin operator:
\begin{align*}
\ket{T_i} &= (\ket{{\uparrow\downarrow}_i} + \ket{{\downarrow\uparrow}_i})/\sqrt{2} \\
\ket{S_i} &= (\ket{{\uparrow\downarrow}_i} - \ket{{\downarrow\uparrow}_i})/\sqrt{2},
\end{align*}
and furthermore transforming the Fock states into eigenstates of the translation operator, labeled by the normalized angular momentum $\ell = 0, \pm 1$:
\begin{equation*}
\ket{x_\ell} = (\ket{x_{i=0}} + e^{i \ell \frac{2\pi}{3}}\ket{x_{i=1}} + e^{i \ell \frac{4\pi}{3}} \ket{x_{i=2}})/\sqrt{3}
\end{equation*}
simplify the single-hole ground state and the first excited states to:
\begin{align*}
\ket{\phi_\text{1h}^g} &\propto \ket{s_{\ell=0}} - \frac{1}{\sqrt{2}}\left(1+\frac{E_g^{1h}}{2t}\right) \ket{d_{\ell=0}} \\
\ket{\phi_\text{1h}^e} &= \ket{x_{\ell=\pm1}}, \quad x = \,\downarrow\downarrow, \, t, \, \uparrow\uparrow
\end{align*}
with eigenenergies:
\begin{equation*}
E_g^\text{1h} = -t + \frac{U}{2} - \sqrt{9t^2+tU+U^2/4} 
\underset{U\gg t}{\sim} -2t -\frac{8 t^2}{U}
\end{equation*}
\begin{equation*}
E_e^\text{1h} = -t
\end{equation*}

The eigenstates and eigenenergies of the single-particle-doped plaquette can be obtained through a particle-hole transformation $\hat{c} \leftrightarrow \hat{c}^\dagger$, $h \leftrightarrow d$, and with $\ket{0}$ the unit-filled state acting as a vacuum state for holes:
\begin{align*}
\ket{\phi_\text{1d}^g} &= \ket{x_{\ell=0}}, \quad x = \,\downarrow\downarrow, \, t, \, \uparrow\uparrow
\\
\ket{\phi_\text{1d}^e} &\propto \ket{s_{\ell=0}} - \frac{1}{\sqrt{2}}\left(1+\frac{E_e^{1h}}{t}\right) \ket{h_{\ell=0}}
\end{align*}
with eigenenergies:
\begin{equation*}
E_g^\text{1d} = -2t
\end{equation*}
\begin{equation*}
E_e^\text{1d} = \frac{1}{2}\left(-t + U - \sqrt{9t^2+tU+U^2/4}\right) 
\underset{U\gg t}{\sim} -t -\frac{2 t^2}{U}
\end{equation*}

The associated spectrum is shown in Fig.~\ref{fig:gs_sim}b. For all positive interactions $U$, the ground state for one particle dopant is one of the three triplet states with angular momentum $\ell = 0$. In contrast, the ground state of a non-frustrated, square plaquette is ferromagnetic only past a critical $U > U_C \approx 18.6$ \cite{yao_myriad_2007}. For one hole dopant, the ground state is a superposition between a singlet and a doublon state with $\ell = 0$. In both cases, the lowest energy gap is equal to the kinetic energy $t$ for $U = +\infty$. The energy of the predominantly singlet states is lowered at finite $U$ by an energy proportional to the spin exchange coupling $J = 4t^2/U$, while triplet states are unaffected.

\subsubsection{Finite Temperature Lanczos Method (FTLM) simulation}\label{subsubsec:ftlm}
We can compute the thermal expectation value $\avg{A}_\beta = \tr{e^{-\beta H} A} / \tr{e^{-\beta H}}$ of arbitrary operators $A$ at inverse temperature $\beta=1/T$ and with Hamiltonian $H$ on finite-sized clusters using the Finite-Temperature Lanczos Method (FTLM) \cite{lanczos1950,prelovsek2017}.
The Lanczos method involves starting from a random state $\ket{r}$ and finding a set of $M$ basis vectors in which the Hamiltonian can be efficiently diagonalized yielding Lanczos approximate eigenvectors $\ket{\psi_i}$ and eigenvalues $\epsilon_i$, for $i=1, ... M$, allowing one to evaluate matrix elements of the form $\avg{r|H^m A|r}$ as long as $m < M$. 
Thermal expectation values can be constructed from these matrix elements as
\begin{equation*}
\begin{split}
    \tr{e^{-\beta H} A} &\approx \sum_{m=0}^{M}{\frac{(-\beta)^m}{m!} \tr{H^m A}} \\
    &\approx \sum_{m=0}^{M}{\frac{(-\beta)^m}{m!} \frac{\text{dim}(H)}{R}\sum_r^R{\avg{r|H^m A|r}} },  
\end{split}
\end{equation*}
where the first approximation comes from truncating the Taylor expansion in $\beta$ to order $M$ leading to error on the order $O(\beta^{M+1})$, while the second approximation comes from using $R < \text{dim}(H)$ states to evaluate the trace which leads to a relative statistical error of the order $O(1/\sqrt{R Z})$ \cite{prelovsek2017} which can be decreased by increased sampling ($Z$ is the partition function, $Z = \tr{e^{-\beta H}}$). 
In all the simulations we use an order $M = 75$ Lanczos decomposition which is typically enough to converge the ground state energy, and use $R=200$ samples in each of the $N_\text{total}$ and $S^z_\text{total}$ symmetry sectors.

We write the Hamiltonian and operators in the Fock basis and work in the $S^z_\text{total}$ conserving sectors as described in \cite{kale2022}. 
In Fig.~\ref{fig:short_range}d and Fig.~\ref{fig:spin}b, we use the $t-J$ Hamiltonian (including the three site terms) \cite{MacDonald1988} with $J=4t^2/U$ to simulate the effects of large interaction strength, including $U/t=\infty$. 
The restricted Hilbert space of the $t-J$ model allows us to reach a system size of $4\times4$ sites with $\text{dim}(H)_\text{max} \sim 2\times 10^6$.
The limited system size introduces some finite size effects which can be seen when comparing simulations on $3\times3$, $4\times3$ and $4\times4$ sites, and also when comparing against other numerical methods such as DQMC on $8\times8$ sites and NLCE simulations.

\subsubsection{Non-interacting calculations}\label{subsubsec:noninteracting}
In a non-interacting Hubbard system, by Wick's theorem, the thermal expectation of any operator written as a product of the fermionic creation and annihilation operators can be evaluated by taking an appropriate sum over all possible contractions of these operators~\cite{AGD}.
Such a sum may be efficiently computed as the determinant of a matrix whose entries are set from the noninteracting Green's function (as discussed, for example, in~\cite{rossi2017}).
We use this technique to compute the correlation functions discussed in Section \ref{subsec:correlators} for arbitrary $\mu,T$ values.
For each $\mu,T$, we compute the non-interacting Green's function for the triangular Hubbard model on a $200\times 200$ mesh in momentum space, using standard formulae~\cite{AGD}, which we convert to real space through a fast Fourier transform. 
In the parameter regimes we access, this choice of mesh is large enough that finite-size effects are negligible.
Since we only compute equal-time correlators we need only store the equal-time Green's function.
This Green's function is then used to fill the matrices whose determinants yield the Wick contractions.

\placefigure[!t]{SG}

\subsubsection{Determinant Quantum Monte Carlo (DQMC) simulation}\label{subsubsec:dqmc}
We use the QUEST package \cite{varney_QUEST_2009} to perform unbiased simulations of the Fermi-Hubbard model on a $8 \times 8$ triangular lattice using the Determinant Quantum Monte Carlo (DQMC) algorithm. DQMC introduces a Hubbard-Stratonovich (HS) transformation to transform the interacting hamiltonian to a non-interacting hamiltonian only quadratic in fermionic operators, but involving a summation over the HS field. For a non-interacting system with $U/t=0$, DQMC becomes exact and computes the non-interacting equal-time Green's function, similar to Section.~\hyperref[subsubsec:noninteracting]{\textit{Non-interacting calculations}}. For interacting system, the summation over HS field is expressed as a classical Monte Carlo problem and can be computed. Thus the operators such as density and correlation functions can be decomposed again using Wick's theorem into the same expressions as non-interacting equal-time Green's functions and computed after performing the HS transformation. 

We added the expression of three-point doublon-spin-spin (hole-spin-spin) correlation function into the QUEST package, which allows us to compute all combinations of $(i,j,k)$ for $\avg{d_i S^z_j S^z_k}$ and $\avg{h_i S^z_j S^z_k}$. The original QUEST package already calculates the doublon density $d$, density $n$ and two-point spin-spin correlation function $\avg{S^{z}S^{z}}$. We can combine these observables to compute the connected correlators. As in \cite{xu_frustration-_2023} we use $5000$ warmup passes and $30000$ measurement passes for each run. At large $U$ or low temperatures, sign problem gets worse we average over $10$ runs initialized with a random seed. Trotterization error also will get worse at large $U$ and we decrease the Trotter step size $td\tau = 0.02$ for $U/t\leq 15$, to $td\tau = 0.01$ for $U/t=15\sim 25$ and $td\tau = 0.005$ for $U/t=25\sim 40$. The values are chosen to make sure the Trotter error is smaller than statistical error. 

\subsubsection{Numerical Linked Cluster Expansion (NLCE)}\label{subsubsec:nlce}
In the NLCE~\cite{nlce1}, an extensive property of the lattice model in the thermodynamic limit is expressed in terms of contributions from all distinct connected (linked) finite clusters, up to a certain size, that can be embedded in the lattice. The method can be summarized as the following series for $P$, the extensive property per site in the thermodynamic limit,
\begin{eqnarray}
P = \sum_c W_P(c),
\label{eq:main}
\end{eqnarray}
where $W_P(c)$ is the contribution of cluster $c$ to the property, calculated recursively starting from $c\equiv$ a single site, according to the inclusion-exclusion principle:
\begin{eqnarray}
\label{eq:exclusion}
W_P(c) = p(c) - \sum_{s\subset c} W_P(s).
\end{eqnarray}
Here, $p(c)$ is the property calculated for cluster $c$ using full diagonalization of the Hamiltonian matrix, and $s$ runs over all subclusters of $c$ (clusters obtained by removing different number of sites from $c$). In practice, clusters that are related by point group symmetry operations of the underlying lattice are grouped together in the above sums. For details of the algorithm, including how to generate clusters and their subclusters for the series on a computer, see Ref.~\cite{nlce2}.\\

We carry out this expansion for both the square lattice~\cite{nlce3} and the triangular lattice Hubbard model to the 9th order, which means we work with clusters of maximum 9 sites. We use the Wynn numerical resummation algorithm~\cite{nlce2} with 3 and 4 cycles of improvement to extend the region of convergence of the series to lower temperatures, typically to $T/t=0.6$ for the triangular lattice around half filling, and use their agreement as an indicator of convergence.

\subsubsection{Density Matrix Renormalization Group (DMRG)}
\label{subsubsec:dmrg}

The ground state DMRG simulations are performed using TeNPy \cite{tenpy} with maximum bond dimension $\chi=2000$ on cylinders of width 4 (Fig.~\ref{fig:spin} and Fig.~\ref{fig:gs_sim}a) and 6 (Fig.~\ref{fig:num_3point}). We perform two-site updates until we reach typical energy convergence of $10^{-6}$ and $10^{-4}$ in the entanglement entropy, followed by one-site updates to further improve the convergence. Observables are averaged over the system size, leading to larger error bars for a few points in the 6-width simulations where our finite bond dimension $\chi\sim 2000$ leads to artificial inhomogeneities.

\subsection{Extended data}
In Fig.~\ref{fig:short_range_suppl}, we show experimental and numerical data for the unnormalized three point correlators for all three interaction strengths of Fig.~\ref{fig:short_range} as a function of doping.

\section*{Data availability}
The datasets generated and analyzed during this study are available from the corresponding author on reasonable request.

\section*{Code availability}
The code used for the analysis are available from the corresponding author on reasonable request.

\begin{dfigure}{SA}
    \noindent
    \includegraphics{"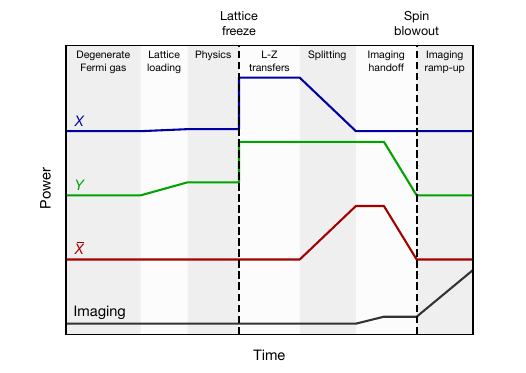"}
    \caption{\textbf{Schematic of experimental sequence.} A degenerate Fermi gas is loaded into a lattice formed by beams $X$ and $Y$ with a linear ramp of the lattice power. The lattice power is quenched to freeze tunneling. Radiofrequency Landau-Zener transfers are used in some shots to change the spin states on singly-occupied sites. Handing off from $X+Y$ to $\bar{X}+Y$ adiabatically doubles the unit cell, converting doubly-occupied sites to pairs of singly-occupied sites. Atoms are handed off to a separate imaging lattice, where a resonant laser is used in some shots to selectively remove one spin state.
    \label{fig:rampfig}}
\end{dfigure}

\begin{dfigure}{SB}
    \noindent
    \includegraphics{"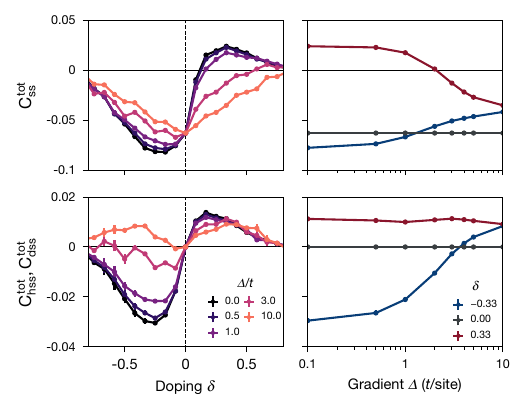"}
    \caption{\textbf{Effect of potential gradients.} Numerical simulation (FTLM) of the nearest-neighbor non-normalized spin-spin and hole-spin-spin (doublon-spin-spin) correlation functions in a $4\times 3$ $t-J$ cluster as a function of doping $\delta$ and gradient strength $\Delta$, at fixed $U/t=30$ and $T/t=0.5$.
    \label{fig:gradient}}
\end{dfigure}

\begin{dfigure*}{SC}
    \noindent
    \includegraphics[width=\linewidth]{"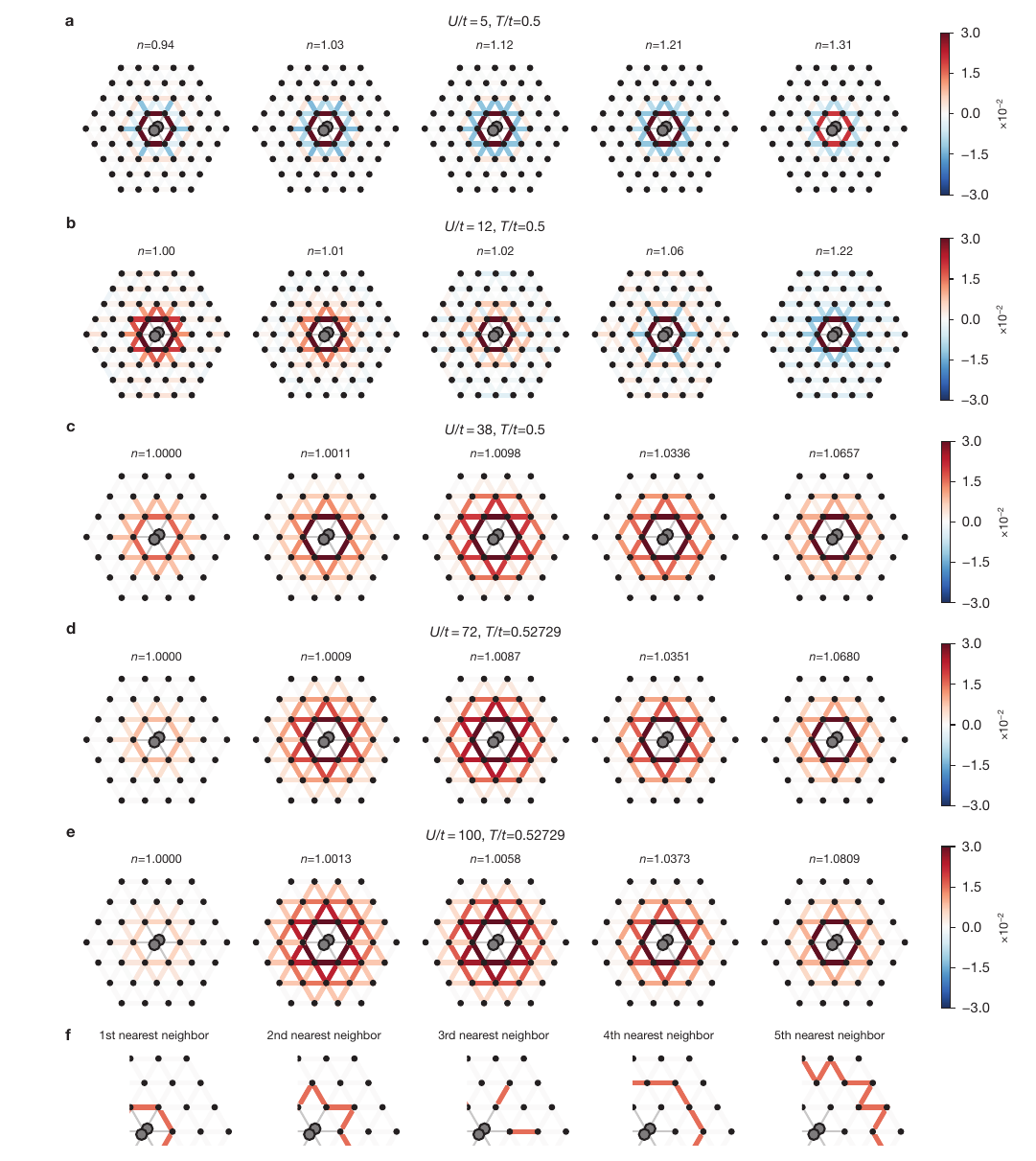"}
    \caption{\textbf{Numerical simulation of doublon-spin-spin correlation map at different densities.}
    We compute the connected doublon-spin-spin correlation function 
    (\textbf{a}) with DQMC at $U/t=5$ and $T/t=0.5$;
    (\textbf{b}) with DQMC at $U/t=12$ and $T/t=0.5$;
    (\textbf{c}) with NLCE at $U/t=38$ and $T/t=0.5$;
    (\textbf{d}) with NLCE at $U/t=72$ and $T/t=0.52729$;
    (\textbf{e}) with NLCE at $U/t=100$ and $T/t=0.52729$.
    (\textbf{f}) Definition of bonds averaged together in NLCE simulations (c-e). Bonds beyond fifth nearest-neighbor are not computed and set to zero in the plot.
    \label{fig:numerical_corrmap}}
\end{dfigure*}

\begin{dfigure}{SD}
    \noindent
    \includegraphics{"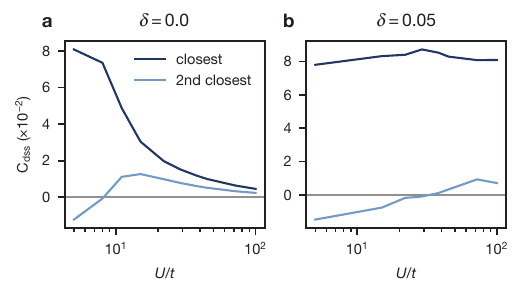"}
    \caption{\textbf{NLCE closest and second-closest doublon-spin-spin correlations.}
    Connected doublon-spin-spin correlator as a function of interaction strength, obtained from NLCE simulations at $T/t = 0.7$ and (\textbf{a}) at half-filling and (\textbf{b}) at particle doping $\delta = 0.05$. See Fig.~\ref{fig:long_range} for a definition of the correlators.
    \label{fig:nlce_fig2}}
\end{dfigure}

\begin{dfigure*}{SE}
    \noindent
    \includegraphics[width=\linewidth]{"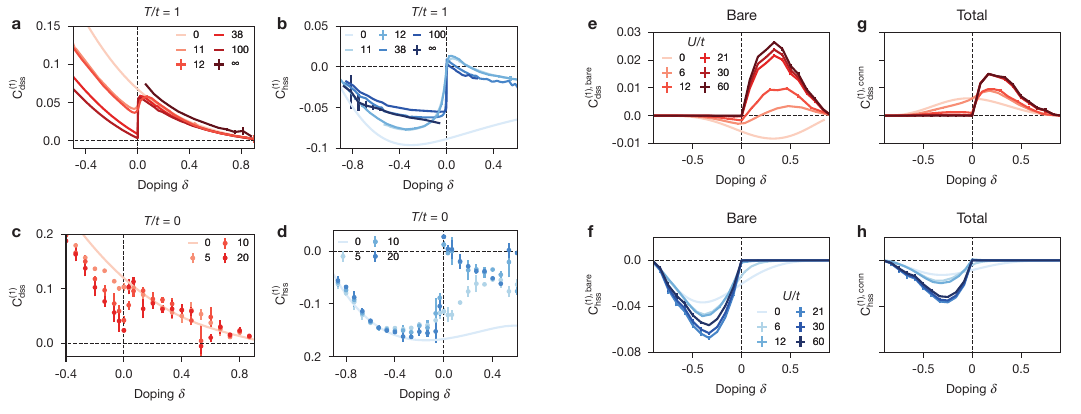"}
    \caption{\textbf{Comparison of numerically computed three-point correlators as a function of doping and interaction strength.}
    (\textbf{a}) -- (\textbf{d}): Comparison between (\textbf{a}), (\textbf{b}) finite-temperature, $T/t = 1$ correlators and (\textbf{c}), (\textbf{d}) ground-state correlators between nearest neighbors, normalized according to Eqs.~\ref{eq:normalized-three-point-correlator}, \ref{eq:connected-three-point-correlator}, \ref{eq:normalized-hss-correlator} and \ref{eq:normalized-dss-correlator-3d} (see Fig.~\ref{fig:short_range}d).
    (\textbf{a}), (\textbf{c}): doublon-spin-spin correlators $C^{(1)}_\text{dss}$, showing an almost universal behavior above half-filling ($\delta > 0$) for the various interaction strengths. (\textbf{b}), (\textbf{d}): hole-spin-spin correlators $C^{(1)}_\text{hss}$.
    The $U/t=0$ numerics are computed using Wick's contraction, $U/t=12$ using DQMC, $U/t= 11, 38, 100$ using NLCE, $U/t=\infty$ using FTLM, and $U/t=5, 10, 20$ using DMRG.
    (\textbf{e}) -- (\textbf{h}): Comparison between (\textbf{e}), (\textbf{f}) bare correlators $C^\text{bare}_{\text{dss}, \text{hss}}$ and (\textbf{g}), (\textbf{h}) non-normalized correlators $C^\text{tot}_{\text{dss}, \text{hss}}$ (as defined in Eq.~\ref{eq:nonnormalized-three-point-correlator} and Fig.~\ref{fig:short_range}c).
    The $U/t=0$ numerics are computed using Wick's contraction, $U/t=6, 12$ using DQMC at $T/t = 0.5$, and $U/t > 20$ using FTLM. The errors in FTLM and DQMC are statistical while in DMRG they indicate the spatial variation of the correlators over the simulated system.
    \label{fig:num_3point}}
\end{dfigure*}

\begin{dfigure}{SF}
    \noindent
    \includegraphics{"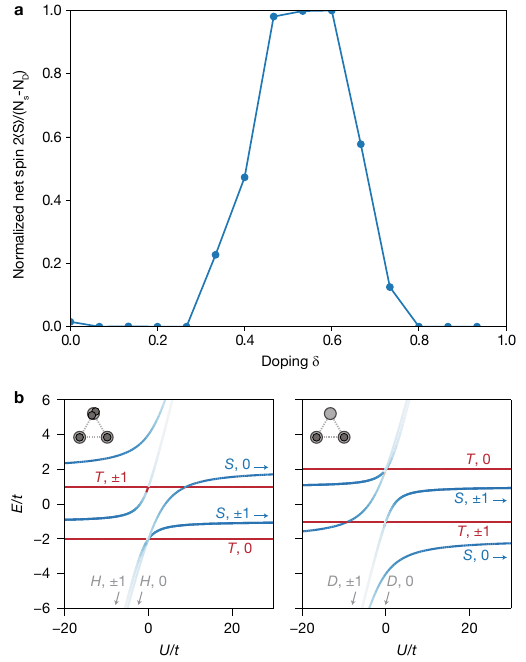"}
    \caption{\textbf{Ferromagnetic state in ground-state simulations.} (\textbf{a}) DMRG simulation of the net total spin $\langle S \rangle$ normalized by maximal spin as a function of doping $\delta$, at $U/t=20$ and zero temperature, showing the emergence of long-range ferromagnetism upon doublon doping. (\textbf{b}) Spectrum of the Hubbard Hamiltonian on a triangular plaquette. Eigenenergies are shown as a function of interaction strength $U/t$ for one particle dopant (left) and one hole dopant (right). Labels show the nature of the state at infinite interaction $U/t = \pm \infty$ ($S$: singlet; $T$: triplet; $H$: one hole; $D$: one doublon) and its angular momentum $\ell = 0, \pm 1$ (see text for definitions). Colors indicate the sign and magnitude of the spin correlations.
    \label{fig:gs_sim}}
\end{dfigure}

\begin{dfigure}{SG}
    \noindent
    \includegraphics{"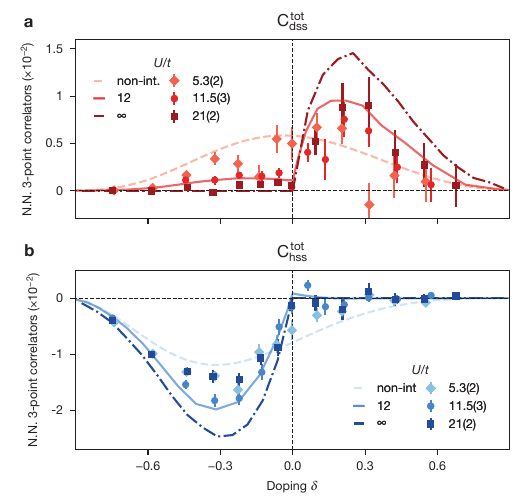"}
    \caption{\textbf{Extended data on doping dependence of 3-point correlators.} (\textbf{a}) Doublon-spin-spin and (\textbf{b}) hole-spin-spin correlation function on a triangular plaquette without normalization factor, see Eq.~\ref{eq:nonnormalized-three-point-correlator} for definition and Fig.~\ref{fig:short_range} for details.
    \label{fig:short_range_suppl}}
\end{dfigure}

\begin{table*}
\centering
\begin{centering}
\begin{tabular}{ | c | c | c | c | c | c | c | c | c | c | c |}
\hline
Dataset & Lattice & $t_{x}$(Hz) & $t_{y}$(Hz) & $t'$(Hz) & Ramp time (ms) & Field (G) & $U/t$ & $T/t$ & Shots & Figures \\
\hline
DS1 & 1 & 380(20) & 360(20) & 370(6) & 160 & 585 & 6(1) & 0.5(2) & 389 & \ref{fig:long_range} \\
\hline
DS2 & 2 & 283(17) & 267(16) & 283(6) & 210 & 610 & 11.5(3) & 0.409(7) & 682 & \ref{fig:long_range} \\
\hline
DS3 & 3 & 183(12) & 172(11) & 173(5) & 330 & 610 & 27(3) & 0.302(5) & 582 & \ref{fig:long_range} \\
\hline
DS4 & 4 & 138(10) & 129(9) & 134(4) & 412 & 610 & 39(2) & 0.306(2) & 1074 & \ref{fig:long_range} \\
\hline
DS5 & 5 & 86(7) & 79(6) & 83(3) & 660 & 610 & 72(11) & 0.30(8) & 1876 & \ref{fig:long_range} \\
\hline
DS6 & 2 & 283(17) & 267(16) & 283(6) & 210 & 610 & 11.5(3) & 0.44(2) & 381 & \ref{fig:short_range} \\
\hline
DS7 & 2 & 283(17) & 267(16) & 283(6) & 210 & 610 & 11.5(3) & 0.65(2) & 259 & \ref{fig:short_range},\ref{fig:spin} \\
\hline
DS8 & 6 & 216(14) & 203(13) & 214(6) & 250 & 610 & 21(2) & 0.31(2) & 199 & \ref{fig:short_range} \\
\hline
DS9 & 6 & 216(14) & 203(13) & 214(6) & 250 & 610 & 21(2) & 0.50(6) & 175 & \ref{fig:short_range},\ref{fig:spin} \\
\hline
DS10 & 2 & 283(17) & 267(16) & 283(6) & 210 & 565 & 5.3(2) & 0.58(8) & 333 & \ref{fig:short_range} \\
\hline
DS11 & 2 & 283(17) & 267(16) & 283(6) & 210 & 565 & 5.3(2) & 0.7(1) & 309 & \ref{fig:short_range},\ref{fig:spin} \\
\hline
DS12 & 2 & 283(17) & 267(16) & 283(6) & 210 & 600 & 9.9(3) & 0.62(4) & 267 & \ref{fig:spin} \\
\hline
DS13 & 3 & 183(12) & 172(11) & 173(5) & 330 & 610 & 27(3) & 0.40(4) & 243 & \ref{fig:spin} \\
\hline
DS14 & 1 & 380(20) & 360(20) & 370(6) & 160 & 607.5 & 8(1) & 0.58(7) & 200 & \ref{fig:spin} \\
\hline
DS15 & 6 & 216(14) & 203(13) & 214(6) & 250 & 595 & 16(1) & 0.40(2) & 298 & \ref{fig:spin} \\
\hline
DS16 & 5 & 86(7) & 79(6) & 83(3) & 660 & 610 & 72(11) & 0.28(8) & 1158 & \ref{fig:spin} \\
\hline
DS17 & 7 & 192(7) & 175(7) & 3.7(2) & 400 & 570 & 9.2(2) & 0.277(7) & 199 & \ref{fig:spin} \\
\hline
DS18 & 7 & 192(7) & 175(7) & 3.7(2) & 400 & 610 & 21.9(4) & 0.6(1) & 195 & \ref{fig:spin} \\
\hline
\end{tabular}
\end{centering}
\caption{\textbf{Summary of experimental datasets.} Note that DS6 is used in both Fig.~\ref{fig:short_range}b and \ref{fig:short_range}d. For each value of $U/t$ in  Fig.~\ref{fig:short_range}d, the hole- and particle-doped curves are obtained from separate datasets (DS6, DS8, and DS10 are hole-doped). Datasets DS11, DS12, and DS13 appear in Fig~\ref{fig:spin}b.}
\label{tab:mega_dataset}
\end{table*}

\begin{table*}
\centering
\begin{tabular}{ | c | c | } 
\hline
Observed outcome & Possible occupation\\
\hline
$d^{(NR)}$ & $d$ \\
\hline
$d^{(R\uparrow)}$ & $d$ \\
\hline
$d^{(R\downarrow)}$ & $d$ \\
\hline
$p^{(NR)}$ & $\uparrow, \downarrow$ \\
\hline
$p^{(R\uparrow)}$ & $\downarrow$ \\
\hline
$p^{(R\downarrow)}$ & $\uparrow$ \\
\hline
$h^{(NR)}$ & $h$ \\
\hline
$h^{(R\uparrow)}$ & $h ,\downarrow$ \\
\hline
$h^{(R\downarrow)}$ & $h ,\uparrow$ \\
\hline
\end{tabular}
\caption{\textbf{Summary of the density resolved and spin-removal imaging technique.} The left column lists all the site-resolved measured outcomes and the right column lists the possible site occupations that map to the same measured outcome.}
\label{tab:density_and_spin_removal_summary}
\end{table*}

\begin{table*}
\centering
\begin{tabular}{ | c | c | c | c |} 
\hline
Dataset & $\kappa_{\rm{maj}}$ ($t/\rm{site}^2$) & $\kappa_{\rm{min}}$ ($t/\rm{site}^2$) \\
\hline
DS1 & 0.036(3) & 0.35(4) \\
\hline
DS6 & 0.085(4) & 0.35(7) \\
\hline
DS7 & 0.060(5) & 0.62(7) \\
\hline
DS10 & 0.068(3) & 0.39(4) \\
\hline
DS11 & 0.053(3) & 0.55(4) \\
\hline
DS12 & 0.043(3) & 0.47(4) \\
\hline
DS14 & 0.036(7) & 0.38(9) \\
\hline
\end{tabular}
\caption{\textbf{Trap curvature in a representative subset of datasets.}}
\label{tab:curvature}
\end{table*}

\end{document}